\documentclass[aps,prb,twocolumn,groupdaddress]{revtex4-2}
\usepackage{graphicx}
\usepackage{float}
\usepackage{amsmath}
\usepackage[T1]{fontenc}
\usepackage{hyperref}

\begin{document}


\title{First Principles Theory of the Pressure Induced Invar Effect in FeNi Alloys}
 

\author{Amanda Ehn, Bj\"orn Alling, Igor A. Abrikosov}
\affiliation{Department of Physics, Chemistry, and Biology (IFM), Link\"oping University, SE-58183 Link\"oping, Sweden}


\date{\today}

\begin{abstract}
The Fe$_{0.64}$Ni$_{0.36}$ alloy exhibits an anomalously low thermal expansion at ambient conditions, an effect that is known as the invar effect. Other Fe$_{x}$Ni$_{1-x}$ alloys do not exhibit this effect at ambient conditions but upon application of pressure even Ni-rich compositions show low thermal expansion, thus called the pressure induced invar effect. We investigate the pressure induced invar effect for Fe$_{x}$Ni$_{1-x}$ for x = 0.64, 0.50, 0.25 by performing a large set of supercell calculations, taking into account noncollinear magnetic states. We observe anomalies in the equation of states for the three compositions. The anomalies coincide with magnetic transitions from a ferromagnetic state at high volumes to a complex magnetic state at lower volumes. Our results can be interpreted in the model of noncollinear magnetism which relates the invar effect to increasing contribution of magnetic entropy with pressure. 
\end{abstract}


\maketitle

\section{Introduction}

In 1897 Ch.E. Guillaume discovered that the face centred cubic (fcc) Fe$_{0.64}$Ni$_{0.36}$ alloy  exhibits an anomalously low thermal expansion for a wide range of temperatures, an effect that is now called the invar effect, and the specific FeNi composition is now called invar. \cite{Guillaume} The effect has been shown to exist in other alloys as well, such as Fe-Pt and Fe-Pd, among others. Besides the anomalously low thermal expansion, other properties such as the magnetisation, elastic constants, and spontaneous volume magnetostriction show anomalous behaviour. \cite{Wassermann,  Nakamura} The effect is utilised in industry for precision tools and instruments. 

It is agreed that an explanation of the invar effect involves a negative contribution to the thermal expansion from the alloys' magnetic properties which compensates for the positive contribution from lattice vibrations. \cite{Matsui} Since the discovery of the invar effect several theories have been suggested to explain it in detail but the topic remains controversial up to date. 

Perhaps the most well-known theory to describe the effect was proposed by Weiss, the so-called 2$\gamma$-state model. \cite{Weiss63} In this model the Fe-atoms can occupy two distinct states: a high spin (HS), high volume and a low spin (LS), low volume state. The HS state has a slightly lower equilibrium energy but a transition from the HS state to the LS state can be thermally induced. At the transition the LS state population is gradually increased while the HS state population decreases. The transition from the high volume state to the low volume state compensates for the unusual thermal expansion. This theory has been supported by several experiments and earlier  \textit{ab initio} calculations \cite{DecrempsNataf04, Rueff, Podgorny}. 

More recently van Schilfgaarde et al. \cite{Schilfgaarde} proposed a new model to explain the anomaly. In their model the anomaly is caused by a continuous transition from a HS ferromagnetic (FM) high volume state to an increasingly complex noncollinear (NC) magnetic state at low volume. The net magnetisation of the system decreases as individual iron spins are canted away from the global magnetisation direction. This noncollinearity results in an anomalous binding energy curve which in a simple model is related to the thermal expansion coefficient via the bulk modulus and Gr\"{u}neisen constant. With the new theory came a newfound interest in the anomaly and several experiments were carried out to investigate its validity. A number of experiments have found results that are consistent with the noncollinear magnetism model \cite{Matsumoto, Kousa, WILDES}, but other studies have found disagreements with the theory. \cite{Decremps, Rueff}

In one experiment stimulated by Ref. \cite{Schilfgaarde} the authors found that the invar effect exists also in Fe$_{x}$Ni$_{1-x}$ alloys with high nickel concentrations upon applying pressure to the material, thus called the pressure induced invar effect. \cite{PressureInd} According to the theory proposed in Ref. \cite{Schilfgaarde} this behaviour could be present in all FeNi alloys as the noncollinearity in the magnetic state is induced by pressure. The anomalous behaviour is related to the transition from the high volume FM to low volume NC state. By applying pressure to Ni-rich Fe$_{x}$Ni$_{1-x}$ alloys the magnetic moments can be induced to form noncollinear magnetic configurations. This pressure induced invar effect has since been reproduced in further experiments. \cite{Matsumoto}

To the best of our knowledge the pressure induced invar effect has not been investigated in detail using \textit{ab initio} methods. We set out to investigate the pressure induced invar effect and the relations between pressure, volume, and magnetic configurations in disordered Fe$_{x}$Ni$_{1-x}$ for x = 0.25, 0.50, 0.64 using \textit{ab initio} theory. We perform calculations on both noncollinear and collinear magnetic states for each composition and volume. We observe a magnetic transition from a ferromagnetic state at high volumes to a complex magnetic state at lower volumes for all compositions. At the corresponding volumes anomalies are observed in the equation of states, although for the most Ni-rich composition this anomaly is very weak. 

\section{Method}

\subsection{Computational details}
The calculations are carried out by the Vienna \textit{ab initio} simulation package (VASP), using the 5.4.4 version. \cite{Kresse1, KRESSE199615, Kresse2} The projector augmented wave (PAW) method is chosen. \cite{PAW} To obtain the 64 atoms supercells we use the special quasi-random structures (SQSs) method. \cite{Sqs} The three supercells have compositions Fe$_{0.25}$Ni$_{0.75}$, Fe$_{0.5}$Ni$_{0.5}$, Fe$_{0.64}$Ni$_{0.36}$, and for the exact number of ion-types see Table \ref{tab} which also contain information about experimental equilibrium lattice parameter and atomic volume.

In order to find the ground state and the corresponding magnetic structure the calculations are initialised with ideal positions of the atoms in a fcc lattice where the ions are allowed to relax around their initial positions. The magnetic moments for the Fe-atoms are initialised with random directions while the Ni-atoms have spins initialised in one direction. During the calculations both the ionic positions and magnetic moments gradually relax toward their most preferred states. Magnetic moments are free to change magnitude and rotate in a noncollinear fashion. To avoid that the calculations are trapped in a local energy minimum far away from the global minimum several calculations are carried out for each composition and volume. For each of these the initialised magnetic structure is a unique random magnetic configuration for Fe moments. In the following sections the lowest energy results for each composition and volume are presented. 

Both NC and collinear magnetism are investigated. In the NC calculations the magnetic moments can evolve in all three dimensions and are thus free to rotate during the optimisation steps. In the collinear case the moments are restricted to a spin 'up' or spin 'down' state. For the collinear calculations the magnetic structure is always initialised in a ferromagnetic state. The magnitudes of the local moments are always free to evolve during the self-consistent cycle. 
 
Calculations are carried out with a 3x3x3 k-point grid generated with the Monkhorst-Pack scheme \cite{MPack}. The Methfessel-Paxton method was used with a smearing of 0.1 eV \cite{Meth}. The energy cut-off is set to 500 eV. The convergence criteria for energy was set to 10$^{-5}$ and for forces 10$^{-4}$. Throughout the work spin-orbit coupling was not taken into account.

\begin{table}[h]
\caption{\label{tab}Experimental lattice parameters and supercell composition information. Column 1 gives the compsition. Column 2 and 3 display the experimental equilibrium lattice parameters  (from Refs.  \cite{experiment,PressureInd}) and corresponding atomic volume for the three compositions. Column 4 and 5 display the number of Fe and Ni-atoms in the supercells, respectively.}
\begin{ruledtabular}
\begin{tabular}{ccccc} 
Composition &  a$_0$ (\AA)  & Atomic volume (\AA$^3$) &	\#Fe & \#Ni  \\
Fe$_{0.25}$Ni$_{0.75}$ & 3.545 &11.14& 16& 48 \\
Fe$_{0.5}$Ni$_{0.5}$ & 3.578 &11.45 &32 & 32\\
 Fe$_{0.64}$Ni$_{0.36}$  & 3.5957& 11.62 &41  & 23\\
\end{tabular}
\end{ruledtabular}
\end{table}

\subsection{Choice of exchange-correlation functional}
The choice of the functional that describe exchange-correlation effects in density functional theory (DFT) requires some discussion. One advantage of  the generalized gradient approximation (GGA) compared to the local-density approximation (LDA) functional is that it typically more accurately reproduces the equilibrium volume for 3d metals. At the same time the GGA overestimates the magnetic moments of the 3d metals and as a result also the magnetic energy. \cite{GGALDA, LDAGGA} Of course, when considering Fe the GGA functionals succeed in reproducing the ground state body centred cubic (bcc) crystal structure of Fe while the LDA generally does not, as discussed in Ref. \cite{conf_igor}. In the same paper it is argued that the reason the stabilisation of the bcc ground state of Fe fails in the LDA is due to a weakening of the magnetic contribution at the equilibrium volume. Meanwhile the GGA corrects the volume and the magnetic contribution, thus savouring the bcc Fe structure. In addition the choice of parameterisation of the GGA functional can produce different results for fcc Fe, as found in Ref. \cite{competition} where different GGA functionals lead to different potential energy landscapes for different magnetic states. 

Previous studies on the magnetic ground state of the invar alloy done with the LDA and GGA functionals yield different magnetic structures in the ground state. GGA functionals lead to a ferromagnetic ordering in the ground state for the invar alloy. \cite{Liot09} In contrast, theoretical studies with the LDA functional have resulted in NC magnetic orderings in the ground state. \cite{Schilfgaarde} Thus the ground state magnetic structure depend on the choice of functional. For this study the LDA functional was chosen as it gives better magnetic properties at a fixed volume \cite{conf_igor} while GGA overstabilises a ferromagnetic state. To address the issue of both functionals finding the wrong equilibrium volume we plot our results in relative change in pressure and volume. To look for trends across functionals the composition x = 0.5 was additionally investigated with the PBE and PW91 functionals. Results from calculations performed with the PBE and PW91 GGA functionals are presented in the Supplementary Material \cite{supplementary}.

\section{Results}

\subsection{Evolution of magnetic structure}
In this section the evolution of magnetic structure for the three studied compositions are presented and analysed.

\subsubsection{Fe$_{0.64}$Ni$_{0.36}$}

\begin{figure}[h]
\centering
\includegraphics[width=.48\textwidth]{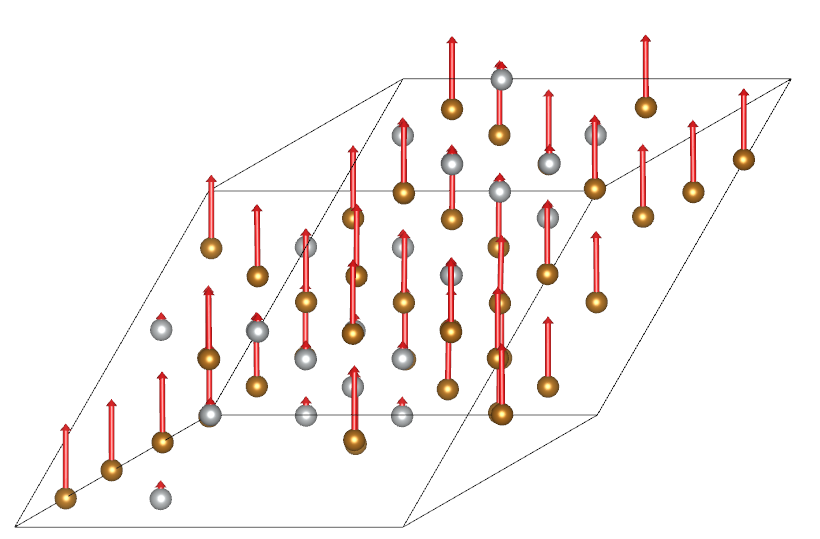}\\
\includegraphics[width=.48\textwidth]{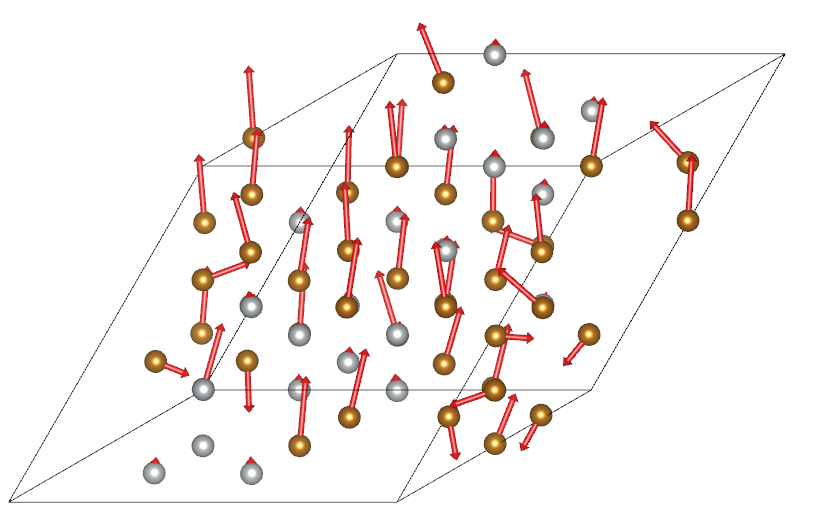}\\
\includegraphics[width=.48\textwidth]{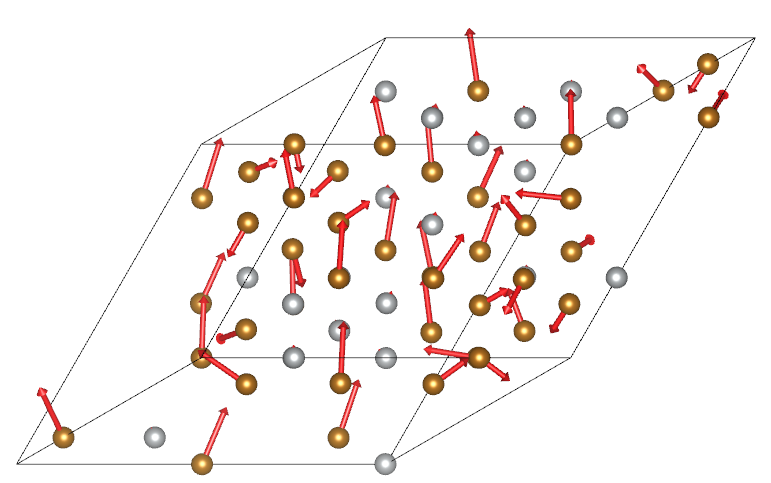}
\caption{Visualisation of magnetic structures of Fe$_{0.64}$Ni$_{0.36}$ at three atomic volumes, from top to bottom: 12.46, 11.57, 11.18 \AA$^3$. Magnetic transition begins at 11.57 \AA$^3$. Gold and silver spheres represent Fe and Ni, respectively.}\label{visual64}
\end{figure}

A visualisation of the magnetic evolution can be seen in Fig. \ref{visual64}, where the magnetic structure is shown at three volumes. The transition from a FM ordering to an increasingly complex structure is seen as the volume is reduced. The transition from a FM state to a noncollinear complex state is continuous and begins at atomic volume 11.57 \AA$^3$, which is 0.5\% smaller than experimental equilibrium atomic volume 11.62 \AA$^3$. These results are in good agreement with the work by van Schilfgaarde et al. \cite{Schilfgaarde} that found a similar transition in the magnetic ordering for the invar composition. For the calculations where the moments are restricted to collinear alignments, a transition with spin-flips begin at 10.95 \AA$^3$. Note that exact values for the transition could be affected by the computational details, e.g. the size of the supercell. However, good agreement between our simulations and those reported in Ref \cite{Schilfgaarde} confirm the reliability of the existence of the transition itself.

\begin{figure}[H]
\centering
\includegraphics[width=.48\textwidth]{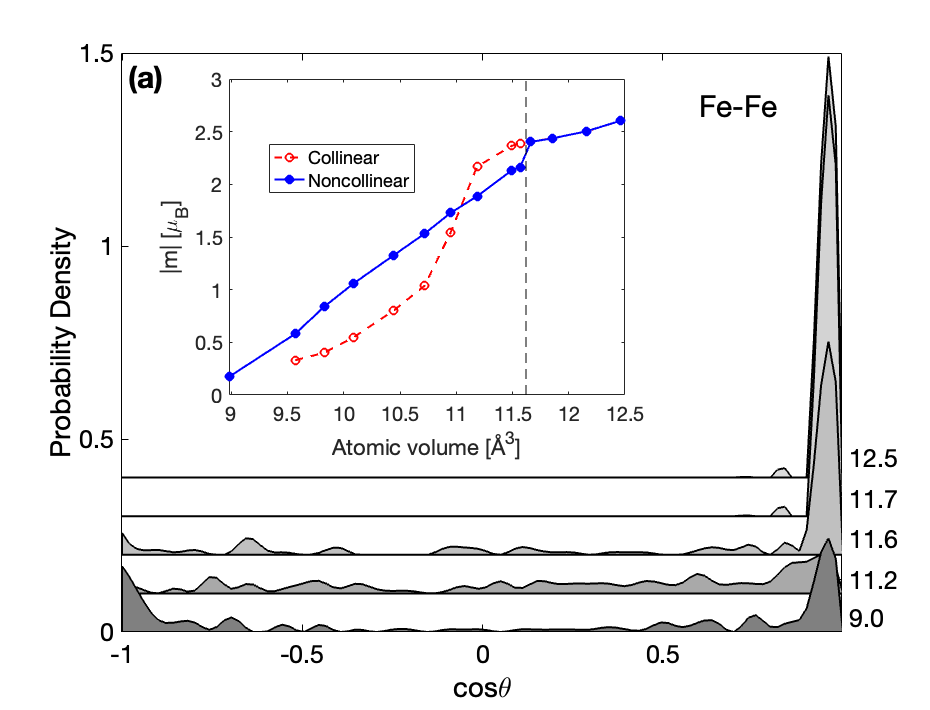}\\
\includegraphics[width=.48\textwidth]{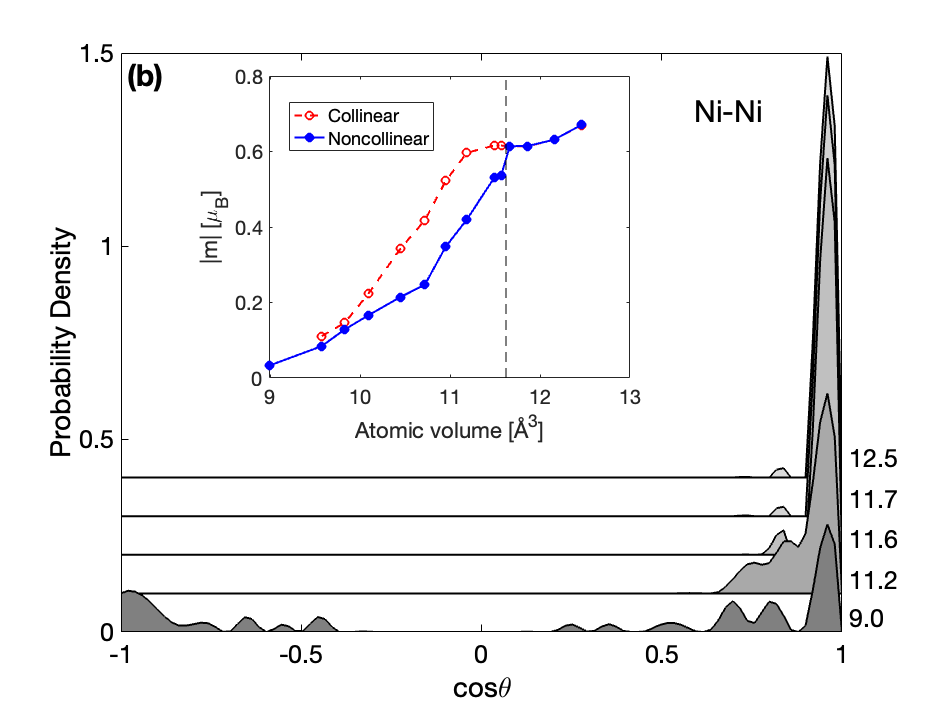}
\caption{Histograms of spin-pair correlation functions for Fe$_{0.64}$Ni$_{0.36}$ between neighbouring Fe-spins (panel \textbf{a}) and Ni-spins (panel \textbf{b}) at several volumes (in \AA$^3$). $\hat{s}_i \cdot \hat{s}_j = cos(\theta)$. The right y-axis gives the atomic volumes in units \AA$^3$. Inserted plot shows the average magnetic moment size for noncollinear and collinear calculations in filled in blue circles and red crosses, respectively. Dashed line is experimental equilibrium volume.}\label{spincorr64}
\end{figure}

\begin{figure}[h!]
\includegraphics[width = 0.48\textwidth]{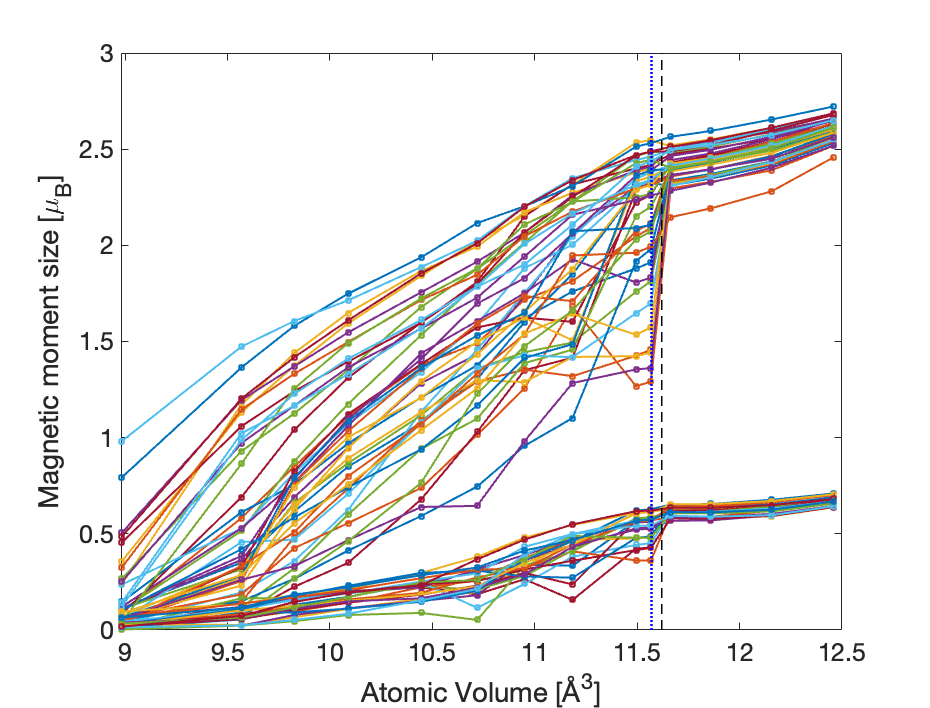}%
\caption{\label{M_64}  Absolute values of local magnetic moments for individual atoms in Fe$_{0.64}$Ni$_{0.36}$, both Fe and Ni, as a function of atomic volume \AA$^3$. Dashed and dotted lines are experimental equilibrium volume and transition volume, respectively.}
\end{figure}

In Fig. \ref{spincorr64} the spin-pair correlation functions for Fe-Fe and Ni-Ni nearest neighbours (NNs) are shown. For the Fe-Fe NNs, we see a transition from a nearly ferromagnetic (FM) state at large volume to a complex state and eventually to a nearly antiferromagnetic (AFM) state. It can be noted that the average Fe moment size is very small at the lowest calculated volume (9.0 \AA$^3$): in Fig. \ref{M_64} individual moments have values $\sim$ 0.25 $\mu_B$. For the Ni-Ni NNs spins remain in the FM state longer upon a reduction of volume, compared to the Fe-Fe pairs, until they too undergo a transition into a complex state, i.e. spin-flips.

When comparing the average magnetic moments for Fe and Ni in the NC and collinear calculations they behave differently. If we start from large volumes and follow the trend in decreasing volume the Fe-atoms show a continuous reduction of the net moment after the magnetic transition from the FM state begins. Once the transition starts the magnetic moment size declines nearly linearly from 2.2 $\mu_B$ at 11.57 \AA$^3$, 1.5 $\mu_B$ at 10.72 \AA$^3$, and 0.8  $\mu_B$ at 9.83 \AA$^3$. The magnetic transition occurs at higher atomic volumes in noncollinear calculations compared to collinear calculations, which can be observed in the inserts in Fig. \ref{spincorr64}. In the collinear calculations the transition start at a smaller volume, and once it starts the average magnetic moment size decreases faster. 

In Fig. \ref{M_64} we show individual magnetic moments sizes from NC calculations, where we observe a sharp drop in the local moment size for some atoms, while for other atoms the local moments decrease is less sharp.

\subsubsection{Fe$_{0.5}$Ni$_{0.5}$}
\begin{figure}[ht]
\centering
\includegraphics[width=.48\textwidth]{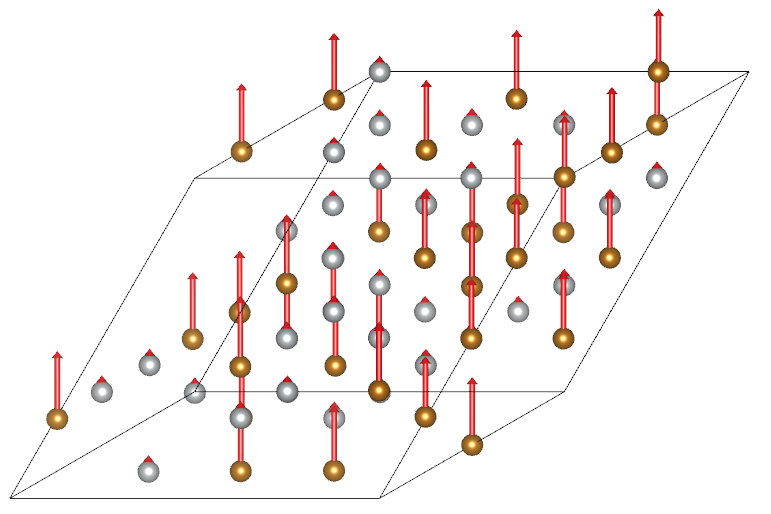}\\
\includegraphics[width=.48\textwidth]{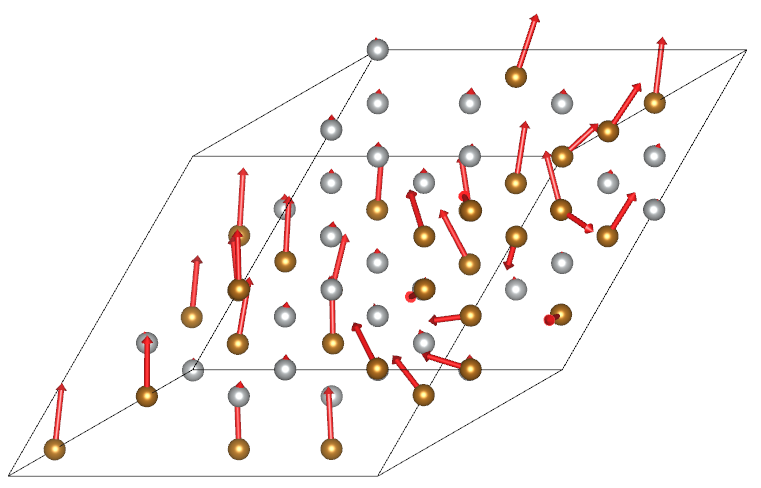}\\
\includegraphics[width=.48\textwidth]{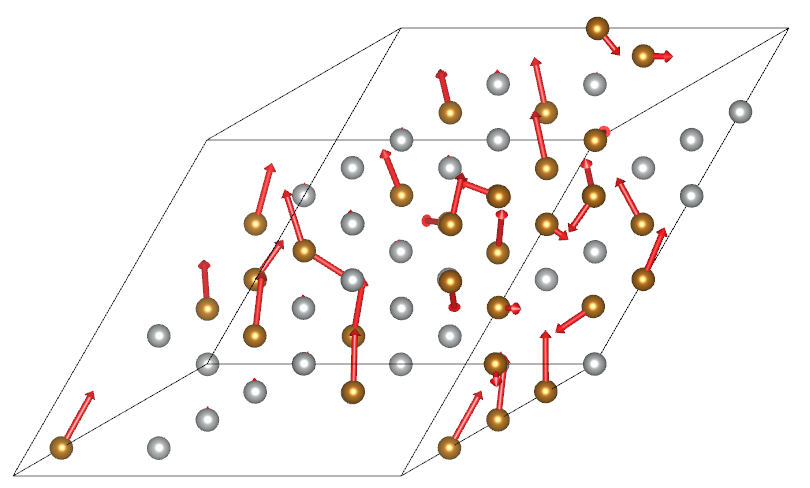}
\caption{Visualisation of magnetic structures of Fe$_{0.5}$Ni$_{0.5}$ at three atomic volumes, from top to bottom: 11.18, 10.95, and 10.54 \AA$^3$. Magnetic transition begins at 10.95 \AA$^3$. Gold and silver spheres represent Fe and Ni, respectively.}\label{vis50}
\end{figure}

A visualisation of the magnetic structure and its evolution can be seen in Fig. \ref{vis50}, where the magnetic structure is shown for three volumes. As the volume is reduced, during the initial stage of the transition the magnetic moments are tilted away from, or in some cases nearly antiparallel, to the global magnetisation direction. Further reduction of the volume increases the complexity in the magnetic structure.

\begin{figure}[h]
\centering
\includegraphics[width=.48\textwidth]{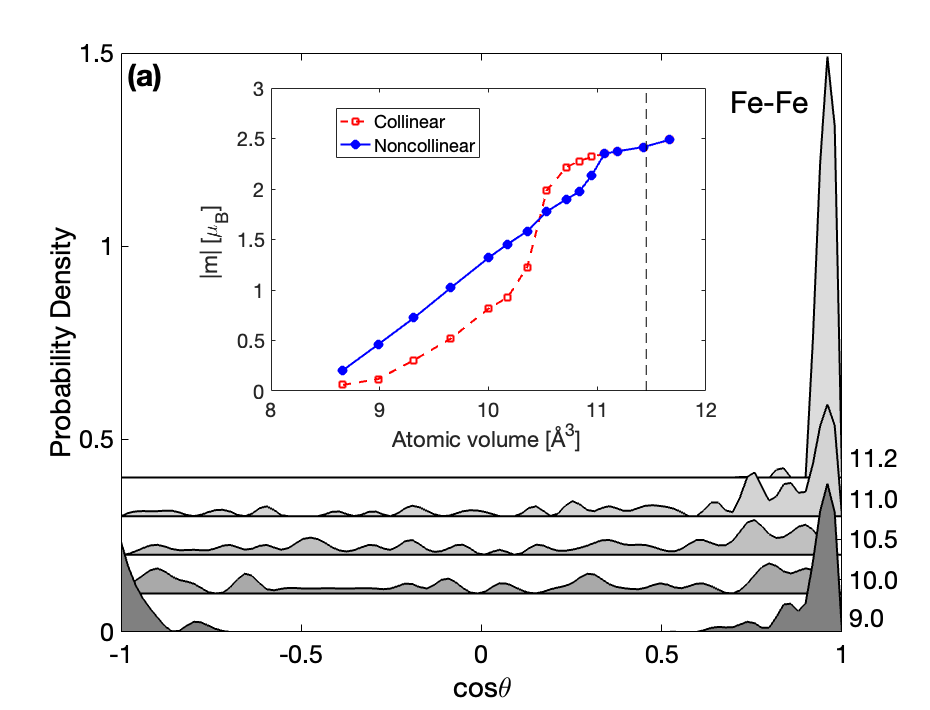}\\
\includegraphics[width=.48\textwidth]{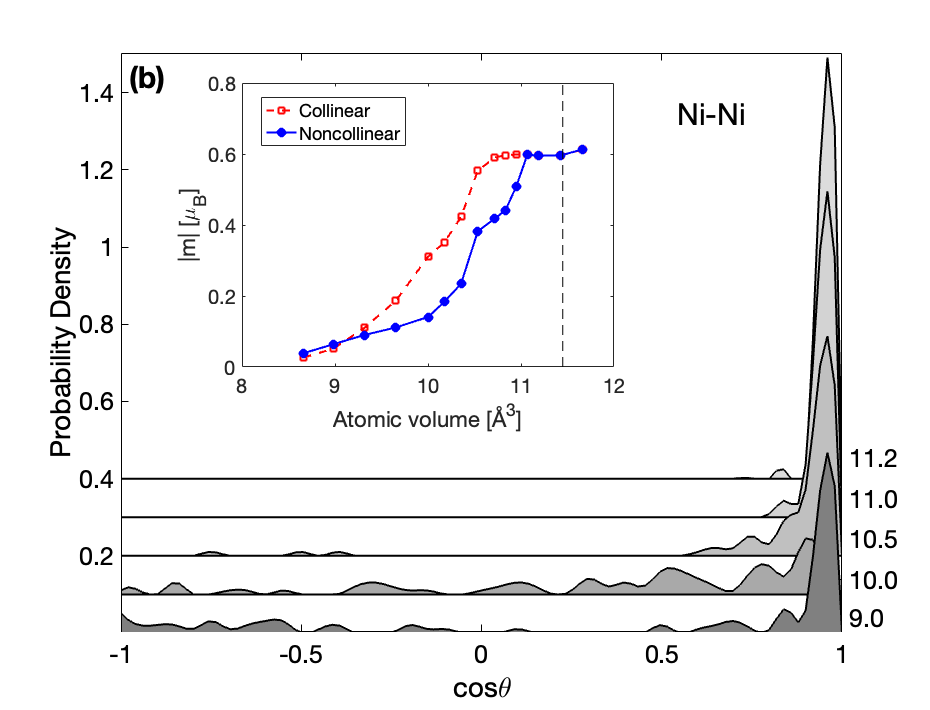}
\caption{Spin-pair correlation functions for Fe$_{0.5}$Ni$_{0.5}$ between neighbouring Fe-spins (panel \textbf{a}) and Ni-spins (panel \textbf{b}) at several volumes (in \AA$^3$). $\hat{s}_i \cdot \hat{s}_j = cos(\theta)$. The right y-axis gives the atomic volumes in units \AA$^3$. Inserted plot shows the average magnetic moment size for noncollinear and collinear calculations in filled in blue circles and red crosses, respectively. Dashed line is experimental equilibrium volume.}\label{spincorr50}
\end{figure}

Similar to the invar composition, we can see that when looking at the spin-pair correlation functions for Fe$_{0.5}$Ni$_{0.5}$ the magnetic state at the largest volume is FM and the transition shows up as a complex state at lower volumes, as is seen in Fig. \ref{spincorr50}. The spin-pair correlation functions behave much as for the invar composition: the Fe-NNs pairs are all parallel at large volumes, but exhibit increasing complexity as the volume reduces until it eventually ends up in an AFM state. For the Ni-Ni spin-pair correlation functions the behaviour is similar to that of the invar composition with deviations from parallel pairs occurring at lower volumes than for the Fe-Fe spin-pairs.

Comparing the result from NC and collinear calculations we observe that the magnetic transition from a FM to a complex state occurs at higher volumes for the NC case, as can be seen in the average magnetisation for Fe and Ni atoms, insets in Fig. \ref{spincorr50}. The transition occurs for the NC case at atomic volume 10.95 \AA$^3$ at which point the average magnetisation for the Fe-moments is reduced. Thus the transition occurs at an atomic volume that is 4.4 \% lower than the experimental equilibrium volume. For the collinear case, the transition begins at 10.36 \AA$^3$, a volume that is 9.6 \% lower than the experimental equilibrium volume. The behaviour at and after the transition differs between the two cases. For the NC case the reduction in moment size is nearly linear, but for the collinear case the decline in average moment size is very rapid. Thus it is observed that the Fe magnetic moments can, on average, retain larger sizes in a complex noncollinear state than when restricted to spin 'up' and 'down' states.

\begin{figure}[h]
\includegraphics[width = 0.48\textwidth]{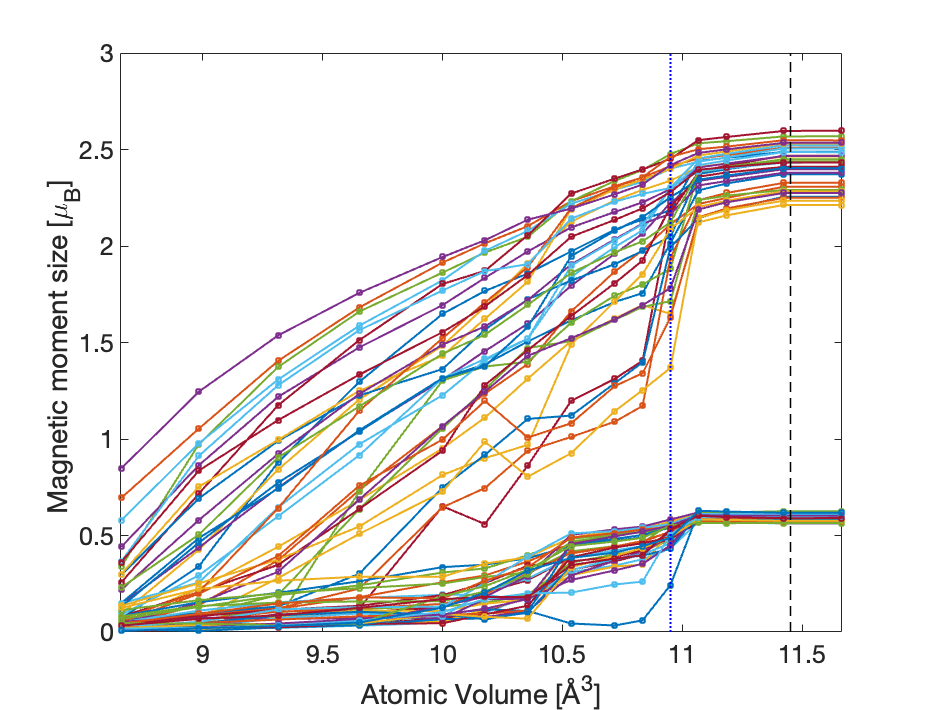}%
\caption{\label{av_M_50_i} Magnetic moment sizes for all individual atoms in Fe$_{0.5}$Ni$_{0.5}$, both Fe and Ni, as a function of atomic volume \AA$^3$.  Dashed and dotted lines are experimental equilibrium volume and transition volume, respectively.}
\end{figure}

For individual atoms' local magnetic moments there are some that decline faster in size compared to the average size, as seen in Fig. \ref{av_M_50_i}. The transition begins at 10.95 \AA$^3$, and here a number of Fe-atoms' magnetic moments decline quickly. 

\subsubsection{Fe$_{0.25}$Ni$_{0.75}$}
In contrast to the two other compositions Fe$_{0.25}$Ni$_{0.75}$ exhibits a smoother variation in average magnetisation and individual local moments. The results from collinear and NC calculations are very close in average magnetisation and local moments.

\begin{figure}[h]
\centering
\includegraphics[width=.48\textwidth]{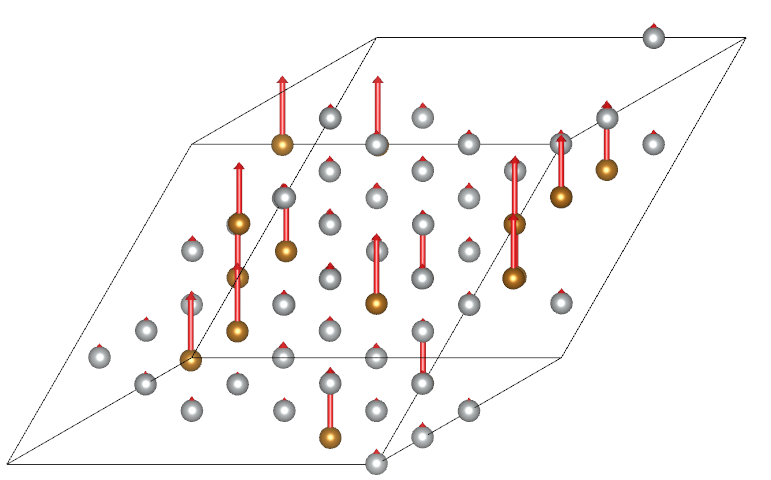}\\
\includegraphics[width=.48\textwidth]{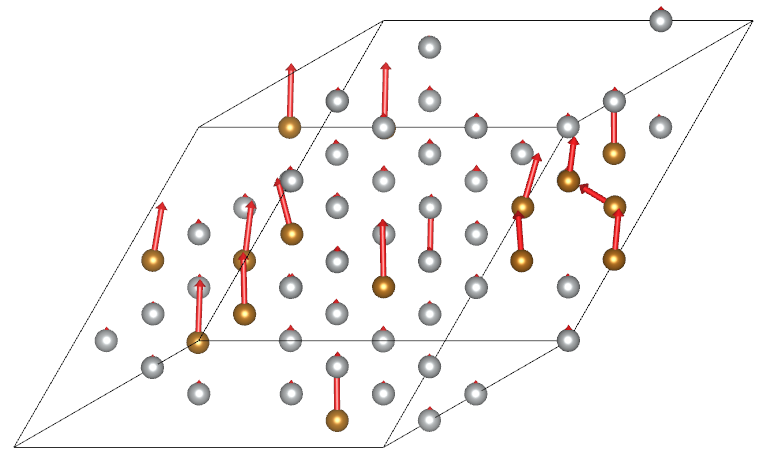}\\
\includegraphics[width=.48\textwidth]{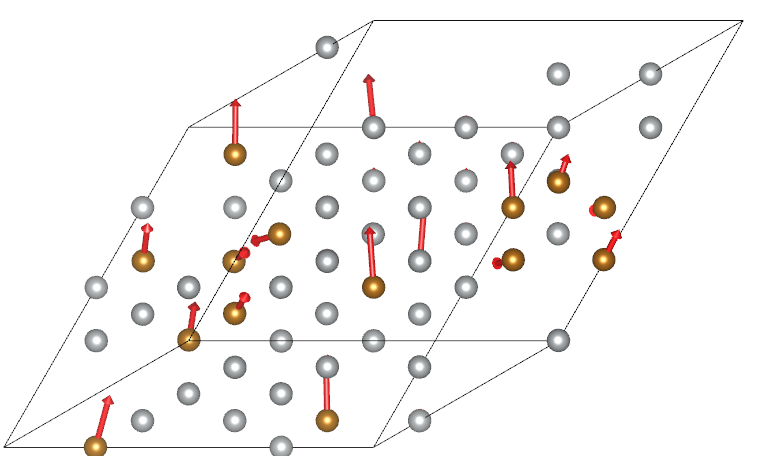}
\caption{Visualisation of the magnetic structures of Fe$_{0.25}$Ni$_{0.75}$ at three atomic volumes, from top to bottom: 10.54, 9.65, 9.31 \AA$^3$. Magnetic transition begins at 9.65 \AA$^3$. Gold and silver spheres represent Fe and Ni, respectively.}\label{vis25}
\end{figure}

Visualisation of the magnetic evolution can be seen in Fig. \ref{vis25} where the magnetic structure is shown for three volumes. At high volumes there is a FM ordering, but upon compression we observe the transition to NC alignments, at least for some atoms. Further reduction in volume increases the complexity in the magnetic structure, though the effect is less pronounced than at higher concentrations of Fe.

\begin{figure}[h]
\centering
\includegraphics[width=.48\textwidth]{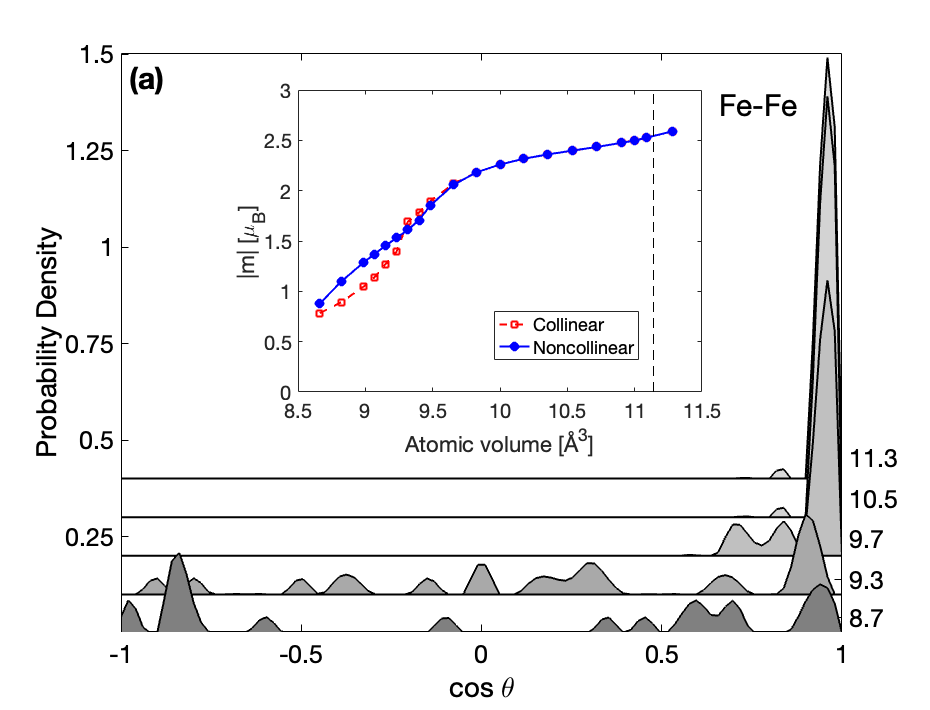}\\
\includegraphics[width=.48\textwidth]{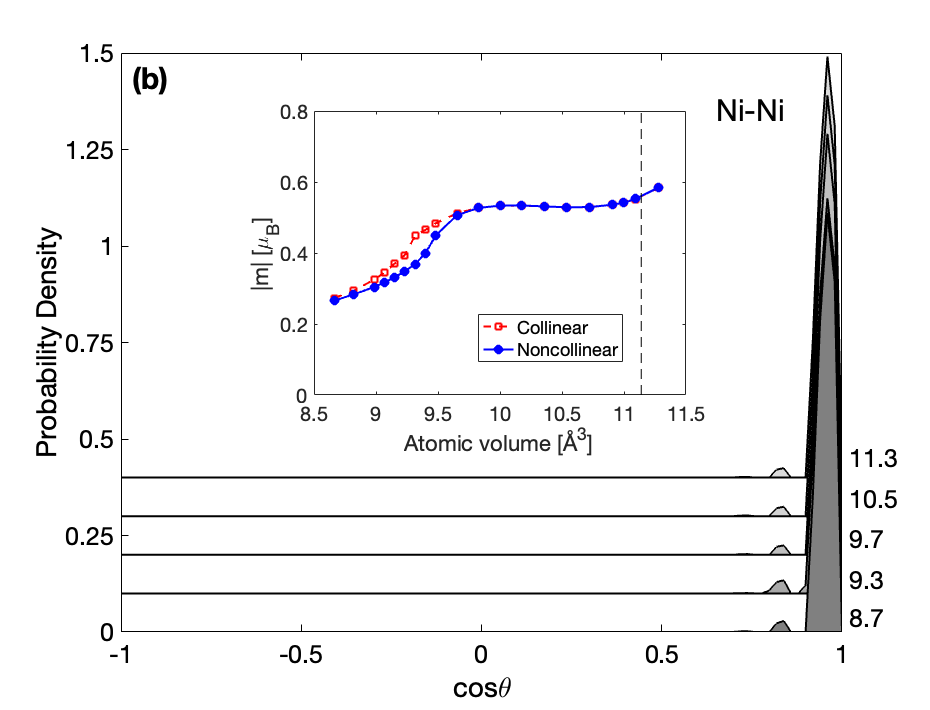}
\caption{Spin-pair correlation functions for Fe$_{0.25}$Ni$_{0.75}$ between neighbouring Fe-spins (panel \textbf{a}) and Ni-spins (panel \textbf{b}) at several volumes (in \AA$^3$). $\hat{s}_i \cdot \hat{s}_j = cos(\theta)$. The right y-axis gives the atomic volumes in units \AA$^3$. Inserted plot shows the average magnetic moment size for noncollinear and collinear calculations in filled in blue circles and red crosses, respectively. Dashed line is experimental equilibrium volume.}\label{spincorr25}
\end{figure}

The spin-pair correlation functions for Fe NNs confirm the NC magnetic structure upon the magnetic transition: the two largest volumes have a FM ordering while the intermediate volume, 9.7 \AA$^3$, has magnetic moments where some pairs are tilted away from each other but still close to the global magnetisation direction. At the lower volumes the pairwise alignments are nearly randomly distributed, as is seen in Fig. \ref{spincorr25}a. Meanwhile the Ni NNs pairs show almost no change in pairwise alignment between the largest and smallest volume, staying parallel at each volume. While the magnetic transition from a ferromagnetic ordering to a complex ordering occurs it happens at atomic volume 9.65 \AA$^3$, 13.3 \% lower than the experimental equilibrium volume. This can be compared to the two other compositions where corresponding numbers are 0.5 \% and 4.4 \% smaller than the experimental equilibrium volume for Fe$_{0.64}$Ni$_{0.36}$ and Fe$_{0.5}$Ni$_{0.5}$, respectively. In calculations with strictly collinear moments the transition with spin-flips begins at 9.15  \AA$^3$.

\begin{figure}[!h]
\includegraphics[width = 0.48\textwidth]{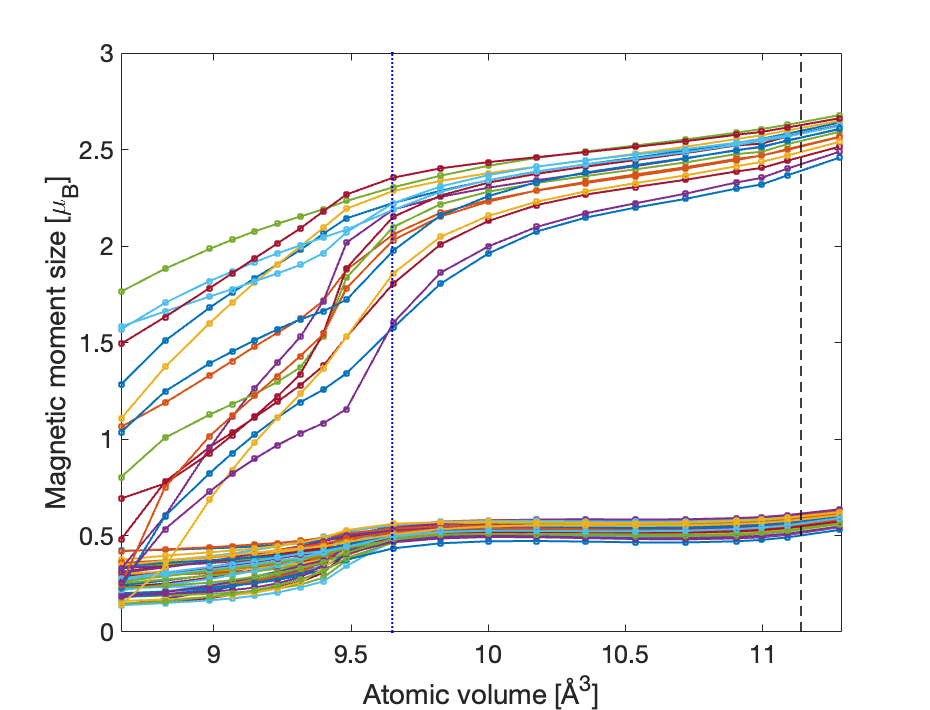}%
\caption{\label{M_25}  Magnetic moment sizes for all individual atoms in Fe$_{0.25}$Ni$_{0.75}$, both Fe and Ni, as a function of atomic volume \AA$^3$.  Dashed and dotted lines are experimental equilibrium volume and transition volume, respectively.}
\end{figure}

The moment sizes for individual atoms are seen in Fig. \ref{M_25}. Interestingly, there are no sharp drops in the magnitudes of local moments, in contrast to the two previous compositions. The magnetic moments slowly decline until volumes of about 9.5 \AA$^3$ at which point some atoms decline quicker as a result of the magnetic transition. 

\subsection{Enthalpy}
\begin{figure}[h]
\centering
\includegraphics[width=0.48\textwidth]{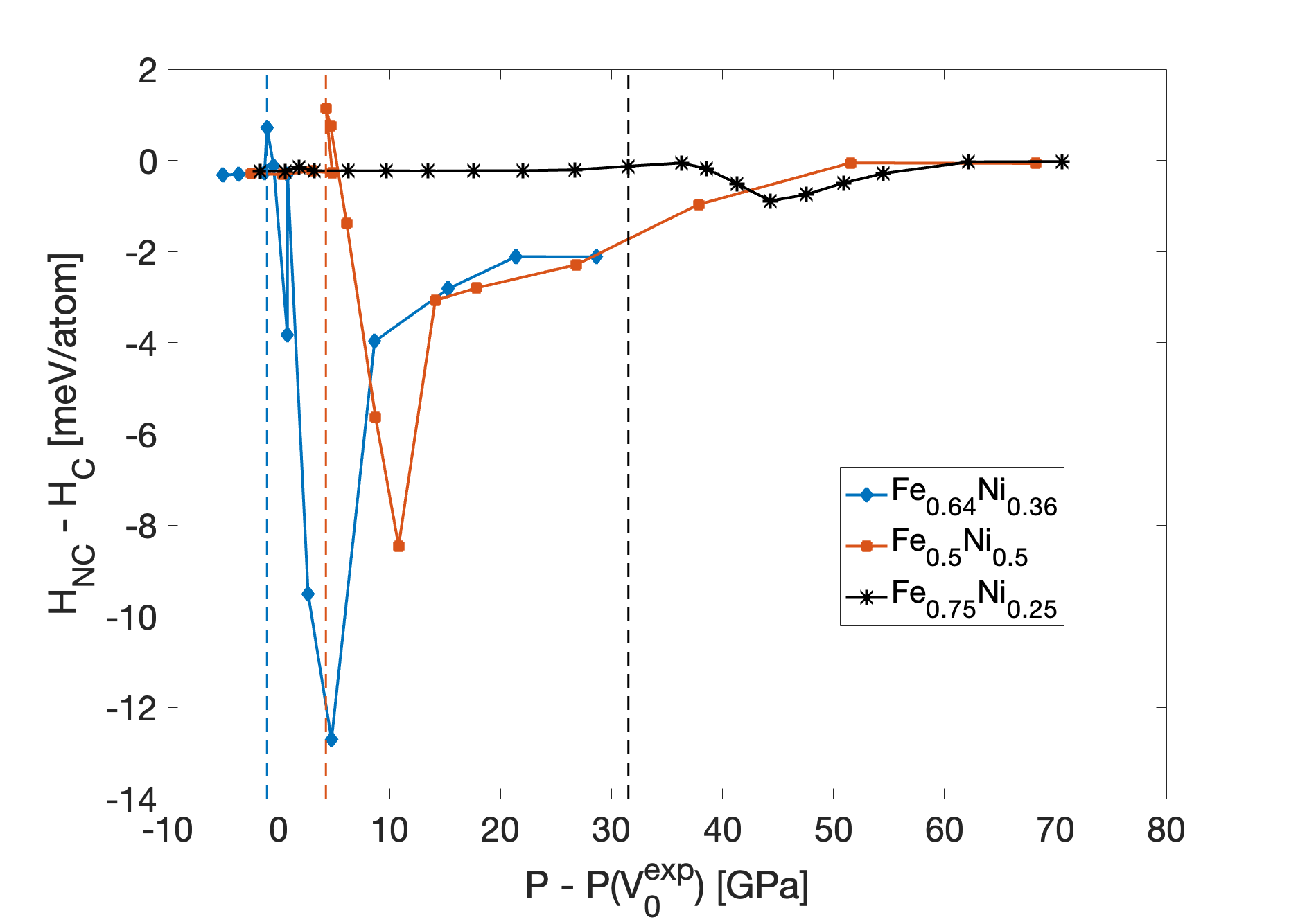}
\caption{\label{deltaH}  Enthalpy differences between noncollinear and collinear calculations for all three compositions. Dashed lines represent the magnetic transitions in NC calculations.}
\end{figure}

The magnetic transitions above are reflected in the enthalpy curves when comparing NC and collinear calculations. This can be seen in Fig. \ref{deltaH} where the difference between the NC and collinear calculations are plotted. Note that $\Delta H$ before the magnetic transition from a FM state to a complex state sets in is within the error bars and most likely caused by the different settings in the calculations. Once the transition begins there are noticeable enthalpy differences for all compositions. The differences become smaller at higher pressures (lower volumes) when the collinear cases exhibit spin-flips. The difference is relatively small though, less than 1 meV per atom, and are thus within the error bars. The difference is likely caused by the different settings in the calculations.

It should be noted that the enthalpy difference between NC and collinear calculations should be 0 for the FM solutions. However, we see a nearly constant shift for all compositions and functionals leading to a negative difference for FM solutions. The settings in the calculations are identical apart from allowing noncollinear and collinear magnetic moments, so this seems to be result from the technical settings in VASP.

\subsection{Equation of state}
\begin{figure}[h]
\centering
\includegraphics[width=0.48\textwidth]{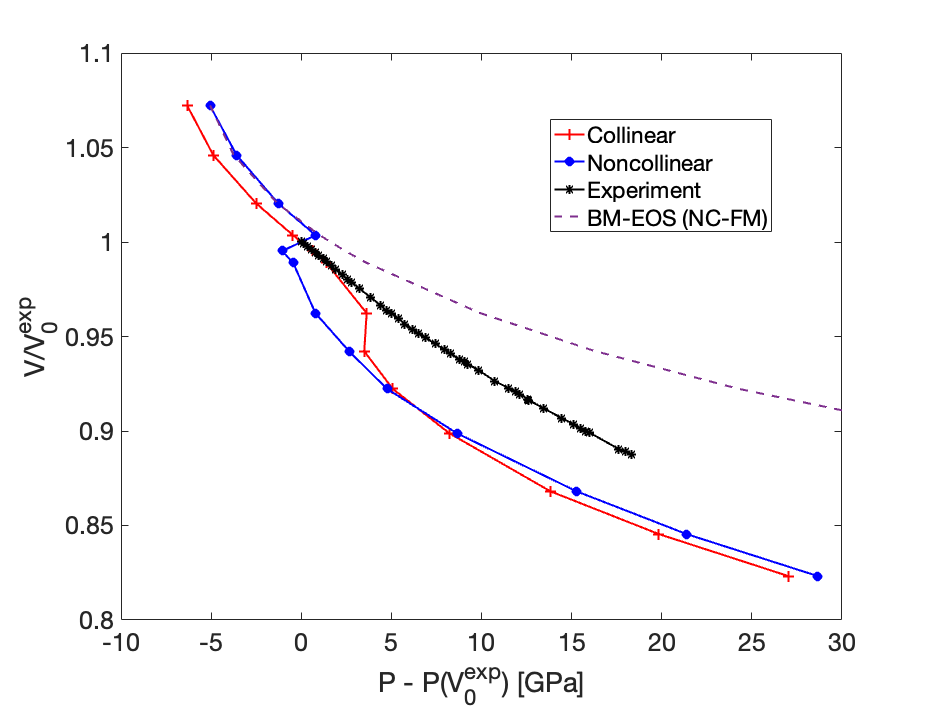}
\caption{\label{PV_64} Equation of state for Fe$_{0.64}$Ni$_{0.36}$ from NC calculations, collinear calculations, and experiment from Ref. \cite{PressureInd}. The dashed line shows the Birch-Murnaghan equation of state fitted to the ferromagnetic datapoints.}
\end{figure}

\begin{figure}[h]
\centering
\includegraphics[width=0.48\textwidth]{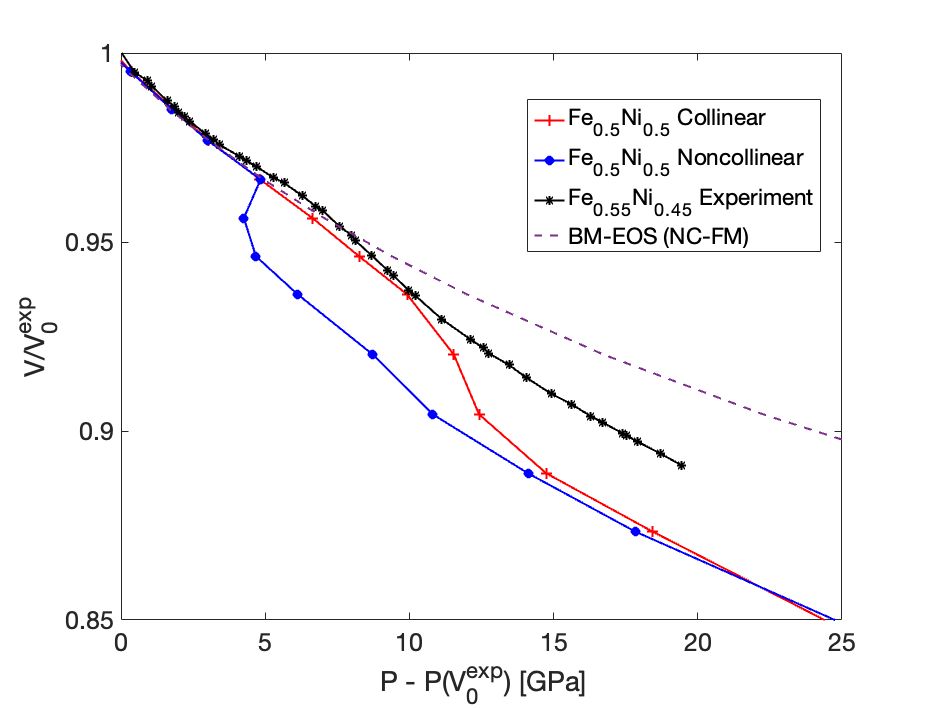}
\caption{\label{PV_50} Equation of state for Fe$_{0.5}$Ni$_{0.5}$ from NC calculations, collinear calculations, and experiment on Fe$_{0.55}$Ni$_{0.45}$  from Ref. \cite{PressureInd}. The dashed line shows the Birch-Murnaghan equation of state fitted to the ferromagnetic datapoints. }
\end{figure}

\begin{figure}[h]
\centering
\includegraphics[width=0.48\textwidth]{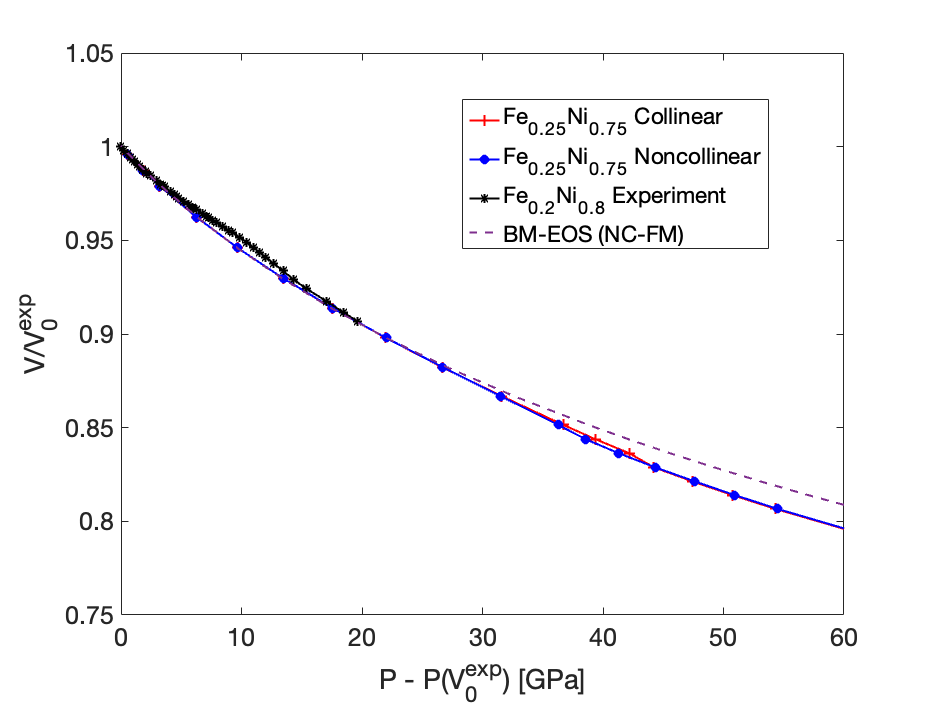}
\caption{\label{PV_25} Equation of state for Fe$_{0.25}$Ni$_{0.75}$ from NC calculations, collinear calculations, and experiment on Fe$_{0.2}$Ni$_{0.8}$  from Ref. \cite{PressureInd}. The dashed line shows the Birch-Murnaghan equation of state fitted to the ferromagnetic datapoints.}
\end{figure}

Next we present the equation of states for the three compositions and compare them with our collinear calculations and  the experimental results from Ref. \cite{PressureInd}. The results are shown in Figs. \ref{PV_64}-\ref{PV_25}. For all compositions from the NC calculations the anomaly appears at volumes corresponding to the onset of the magnetic transition from a FM state to a complex state, although for Fe$_{0.25}$Ni$_{0.75}$ the anomaly is extremely weak. As both NC and collinear calculation reproduce this magnetic transition, albeit at different volumes, the anomalies in the equation of state (EOS) appear at separate volumes. The Birch-Murnaghan equation of state is fitted to the ferromagnetic results in each composition (i.e. to datapoints corresponding to volumes higher than the magnetic transition volume) and shown as a dotted line. For easier comparison with results in previous sections, Table \ref{vol-proc1} relates the volumes and magnetic transition volumes to experimental equilibrium volumes in the studied Fe-Ni alloys.

\begin{table}[h]
 \caption{\label{vol-proc1} Magnetic transition volumes from calculations and $V/V_0^{exp}$ for the studied compositions of fcc Fe-Ni alloys.}
 \begin{ruledtabular}
 \begin{tabular}{ccc}
 Composition  & $V_{FM \rightarrow NC}$ (\AA$^3$) & $V_{FM \rightarrow NC}/V_0^{exp}$\\ \hline 
 Fe$_{0.64}$Ni$_{0.36}$ &  11.57 & 0.996 \\
 Fe$_{0.5}$Ni$_{0.5}$ &  10.95 & 0.956 \\
 Fe$_{0.25}$Ni$_{0.75}$ &  9.65 & 0.866 \\
 \end{tabular}
 \end{ruledtabular}
 \end{table}

\section{Discussion}
As was pointed out earlier \cite{conf_igor, competition} and additionally underlined in Sec. II.B the existing approximations for the exchange-correlation functionals within DFT might not be sufficient for a quantitatively accurate and detailed description of Fe-Ni invar alloys. Indeed, comparing results calculated for the fcc Fe$_{0.5}$Ni$_{0.5}$ alloy (Sec. III) with GGA calculations (see Supplementary Material and more detailed discussion below) one sees their significant dependence on the chosen computational scheme. However, many essential features of the concentration and pressure dependences of magnetic, thermodynamic and structural properties calculated for the fcc Fe-Ni system in this work are general, and therefore we start our discussion with the observations that do not depend on the details of the methodology.  

Most importantly, our calculations clearly demonstrate the existence of a magnetic transition from the ferromagnetic state at large volumes to increasingly NC configurations upon volume decrease. This effect is observed for all the alloy compositions considered in this study, not just at invar composition. However, in Fe$_{0.64}$Ni$_{0.36}$ invar the magnetic transition occurs in an immediate vicinity of the experimental equilibrium volume (at the LDA level of theory), while in alloys with higher Ni concentrations it is induced by pressure, in agreement with experimental observations of the pressure induced invar effect. \cite{PressureInd} The transition is initiated at Fe atoms, implying that the Fe atoms' magnetic moments are driving the transition, and thus they are responsible for the anomalous behaviour of properties of the alloys. Upon the transition from the FM to the NC state both longitudinal and transverse degrees of freedom are affected: Fe local moments shrink Figs. \ref{M_64}, \ref{av_M_50_i}, \ref{M_25}) and become canted (Figs. \ref{visual64}, \ref{vis50}, \ref{vis25}). On the other hand, local moments at Ni preserve their FM alignment to much lower volumes in comparison to Fe (Figs. \ref{spincorr64}, \ref{spincorr50}, \ref{spincorr25}). This is so, because the Ni moments are known to be more itinerant, and therefore they are oriented along the net magnetisation direction of the system. In fact, this observation agrees with a recent experimental observation of very high stability of the ferromagnetic state in fcc Ni, with the Curie temperature above the room temperature at least up to 260 GPa. \cite{PhysRevLett.111.157601}

Comparison of the enthalpies between NC and collinear calculations (Fig. \ref{deltaH}) clearly favour the NC model of the pressure induced magnetic phase transition. Interestingly, for all the studied compositions our calculations predict that upon further increase of pressure the NC state evolves towards the LS nearly AFM state at the lowest volumes considered in this study (see Figs. \ref{spincorr64}, \ref{spincorr50}, \ref{spincorr25}). Though the stability of the LS AFM states with respect to more NC states at the highest considered pressure is at the limit of the accuracy of our calculations (see Supplementary Material Figs. 1-3 and Tables I-III), the fact that the ''starting'' and the ''end'' points of the magnetic transition are the HS FM and LS AFM states, respectively, is seen at all the compositions. Moreover, it does not depend on the approximation for the DFT exchange-correlation functional (see Supplementary Material Figs. 4 and 5, and Tables IV and V). This observation provides a link between the classic Weiss 2$\gamma$-state model of the invar effect \cite{Weiss63} and the more recent NC picture \cite{Schilfgaarde}, confirmed also in this study. 

It is clearly seen that the magnetic phase transitions lead to well defined peculiarities at the EOS at all the considered compositions, Figs. \ref{PV_64}-\ref{PV_25}, in qualitative agreement with experiment. \cite{PressureInd} A discussion of the quantitative agreement is more challenging due to the limited accuracy of the existing exchange-correlation functionals. Let us reiterate an important point from Sec. IIB: to reduce the uncertainty due to computational details, we plot the data in Figs.  \ref{PV_64}-\ref{PV_25} as $V/V_0^{exp}$ vs $P-P(V_0^{exp})$, where the latter pressure, $P(V_0^{exp}$), is the calculated pressure at the experimental volume (obviously, it is 0 for the experimental data). 

Starting with simulations for Fe$_{0.64}$Ni$_{0.36}$ invar composition in the LDA, the NC calculations predict that the deviation from the EOS fitted to the FM data (dotted line in Fig. \ref{PV_64}) occurs when the magnetic transition begins at volume 11.57 \AA$^3$, corresponding to $V/V_0^{exp} = 0.995$.  A similar deviation is observed in the collinear calculations when the magnetic moments begin to flip, at volume 10.95 \AA$^3$ ($V/V_0^{exp} = 0.942 $). As has been pointed out already in Ref. \cite{PressureInd}, the peculiarity at the experimental curve is hardly visible, because the FM branch of the EOS corresponds to a negative pressure, and therefore cannot be measured. However, we note good agreement between the curvature of the experimental calculated EOS above the transition, and a significant deviation of the curvature from the extrapolated EOS of the HS FM state.  In fact, it is reasonable to assume that experimentally the Fe$_{0.64}$Ni$_{0.36}$ invar alloy is in the very vicinity of the magnetic phase transition and is nearly ferromagnetic at its equilibrium volume. This is supported by a high experimental value of its net magnetisation, which is (almost) at the Slater-Pauling curve when measured at temperature T=77K \cite{landolt}, as well as by an observation from neutron scattering experiments with polarisation analysis that the magnetic moments must be collinear on atomic length scales at ambient pressure \cite{WILDES}. With this assumption, we argue that the position of the peculiarity at the EOS calculated in the NC model using LDA is in good agreement with experiment. 

In this respect, the simulations for the equiatomic alloy are more informative, as both theory and experiment capture the HS FM state at low positive pressure, and therefore see the whole transition. From the experiment one sees that there is an anomaly at the curve between 5 and 9 GPa. Pronounced anomalies are also observed in our \textit{ab inito} results. In the NC calculations the deviation occurs when the magnetic transitions begins at atomic volume 10.95 \AA$^3$ ($V/V_0^{exp} = 0.956 $). A similar deviation is observed in the collinear calculations when the magnetic moments begin to flip, but at significantly lower atomic volume 10.36 \AA$^3$ ($V/V_0^{exp} = 0.904$ ). In fact, the equation of state for the NC calculations locates the EOS anomaly at relative volumes and pressure that are in very good agreement with the experiment, though the experimental results are for a slightly lower Ni-concentration, $Fe_{0.55}Ni_{0.45}$. While position of the anomaly and the curvatures of the both HS FM branch above the transition and the LS AFM branch below $P-P(V_0^{exp})$ ~ 10 GPa are well reproduced, the volume collapse upon the transition is significantly overestimated. In fact, similar effect is seen in LDA calculations for Fe$_{0.64}$Ni$_{0.36}$ invar alloy following the assumption that it is (nearly) ferromagnetic at equilibrium volume, as discussed above. 

To investigate sensitivity of the volume collapse to computational details, the evolution of magnetic structure and the EOS of the fcc Fe$_{0.5}$Ni$_{0.5}$ alloy were simulated using two different forms of the GGA, the PBE and PW91. The results are presented in Figs. 6-15 in the Supplementary Material. Qualitatively, they are quite similar to the LDA results. Looking at the EOS calculated using GGA, Figs. 9 and 14, we also observe the peculiarities induced by the magnetic phase transition. While $V/V_0^{exp}$ at the transition from FM to NC state is significantly lower than experimental value, especially for the PBE functional form of the GGA, the magnitude of the volume collapse upon the transition is very well reproduced, in contrast to the LDA EOS. This observation strengthens our confidence in relating experimentally observed peculiarity of the EOS to the pressure induced FM-to-NC transition of the magnetic structure of the alloy.  

Moreover, we would like to point out that the original interpretation of the experimental results in Ref. \cite{PressureInd} was that the anomaly at each composition should be interpreted as a 'bump' on the (conventional) P-V curves. Our results reproduce anomalies in both noncollinear and collinear calculations but offer an alternative interpretation. We believe that the EOS should be considered as consisting of the two branches, corresponding to the HS FM state at high volumes and LS AFM (probably followed by the non-magnetic) state at low volumes smoothly connected to each other in the (relatively narrow) transition region. Such an interpretation also provides a link between the 2$\gamma$-state model of Weiss \cite{Weiss63} and the NC picture. \cite{Schilfgaarde} 

In addition, we observe differences in the magnetic behaviour of different Fe atoms, Figs. \ref{M_64}, \ref{av_M_50_i}, \ref{M_25}. These differences are due to different chemical local environments and connects to the main focus of the theories of F. Liot that observed Fe spin-flips in Fe-rich local environments in collinear calculations. \cite{Liot09, competition} However, our noncollinear calculations show that many of the Fe moments do participate in noncollinear rotations. Typically in Fe-rich local environments but not only those in the most Fe-rich environments. Further investigations, using larger supercells, are needed to fully quantify the precise local environment where magnetic transitions away from the FM state happens at the largest volume. Thus our theory also provides a link between previous works focusing mainly on the global composition, \cite{Ruban1, Ruban2, Johnson1, Johnson2, Crisan, Akai, CPA} and those with full focus on the local environment effects. \cite{Liot09, competition} Furthermore, it may be useful for a thermodynamic description of Fe-Ni invar alloys, e.g. using methods of computational thermodynamics, like CALPHAD.   

Turning now to a comparison between the experimental EOS for the Ni-rich fcc Fe$_{0.20}$Ni$_{0.80}$ alloy and the calculated EOS for the Fe$_{0.25}$Ni$_{0.75}$ alloy shown in Fig. \ref{PV_25}, one sees for the former the behaviour of the P-V curve is anomalous between 9 and 14 GPa. In contrast to alloys with lower Ni concentrations, one does not see pronounced anomalies in the \textit{ab inito} results. However, according to our calculations the magnetic transitions in NC calculations begins at volume 9.65 \AA$^3$ ($V/V_0^{exp} = 0.866 $), and one sees that at the transition the calculated EOS starts to deviate significantly from the curve obtained by the extrapolation of the FM EOS (the dotted line in Fig. \ref{PV_25}).  The transition volume is clearly underestimated for the Fe$_{0.25}$Ni$_{0.75}$ alloy. Note however, that the spin flips in the collinear calculations start to occur at even lower volume 9.15 \AA$^3$ ($V/V_0^{exp} = 0.821$ ). On the other hand, the volume collapse is reproduced much better in comparison to simulations for the alloys with lower Ni concentration already at the LDA level of the DFT. One possible reason for the underestimation of the transition volume in Ni-rich alloys is that the transition is initiated at Fe atoms in Fe-rich chemical environments, as discussed above. For the Ni-rich alloys they could not be well represented in our relatively small supercell. On the contrary, they are present in real macroscopic alloys, and act as nuclei that initiate the transition, most probably at higher volumes. 

Summarising the discussion of the EOS, we conclude that our results provide convincing explanation of the pressure induced peculiarities observed at the EOS of fcc Fe-Ni alloys in experiment \cite{PressureInd} and identify it with the FM-to-NC magnetic transitions upon the compression of the alloys. In its own turn, the experiment unambiguously couples the EOS peculiarities to the very low thermal expansion measured for the alloys at the pressures corresponding to the peculiarities (see Fig. 4 of Ref. \cite{PressureInd}). In this paper we do not aim to calculate the thermal expansion explicitly, as this requires highly time-consuming \textit{ab initio} molecular dynamics simulations coupled to the simulations of the spin dynamics, for instance using the methodology of \cite{PhysRevLett.121.125902}. Such simulations are beyond the scope of the present study and would be reported elsewhere. However, a qualitative picture could be captured based on simple thermodynamic arguments put forward by Abrikosov et al. in \cite{competition}. Let us connect the temperature dependence of volume V to the pressure dependence of entropy S  via Maxwell relations: 

\begin{equation}
\label{EQU}
\left(\frac{\partial{V}} {\partial{T}}\right)\biggr\rvert_P =  -\left(\frac{\partial{S}} {\partial{P}}\right)\biggr\rvert_T
\end{equation}

Clearly, the increasing complexity and frustration of the alloys magnetic structure, both in longitudinal and transverse degrees of freedom, which is in fact maximal in the transition region (Figs. \ref{visual64}, \ref{M_64}, \ref{vis50}, \ref{av_M_50_i}, \ref{vis25}, \ref{M_25}) increases entropy making the magnetic contribution to the term in the right-hand size of Eq. (\ref{EQU}) positive. The corresponding contribution to the temperature dependent volume expansion in the left-hand side of Eq.  (\ref{EQU}) is therefore negative, compensating for the positive contribution due to thermal lattice disorder. In fact, recent calculations by Heine et al. \cite{phonon} have demonstrated that for the fcc Fe$_{0.65}$Ni$_{0.35}$ invar alloy at ambient pressure the contributions due to the magnetic and vibrational disorder cancel each other exactly. Our study underlines that the magnetic complexity at the HS FM-to-NC transition increases greatly in comparison to non-invar compositions and pressures.  

\section{Conclusions}
We have carried out a large set of supercell calculations to investigate the pressure induced invar effect in the fcc Fe$_x$Ni$_{1-x}$ alloys for three compositions of Fe: 64, 50 and 25 at\%. A transition from the high-spin ferromagnetic ordering at large volumes to a complex non-collinear state at lower volumes is detected in all three alloys considered in our study. At the transition volumes we observe significant anomalies in the equation of state. These results can be interpreted within the model of noncollinear magnetism for the invar effect, in which the increasing magnetic complexity in the region of the HS FM-to-NC transition leads to a positive sign of the pressure dependence of the magnetic entropy, which in its own turn is coupled to anomalously large negative magnetic contribution to the temperature dependence of volume of invar alloys via Maxwell relation, Eq. (\ref{EQU}). For the invar composition the HS FM-to-NC transition occurs at ambient conditions, but for the more Ni-rich compositions it can be induced by external pressure. Thus our first-principles theory provides convincing qualitative explanation for the pressure induced invar effect, discovered in Ref. \cite{PressureInd}. 

\section*{Data availability}
All data can be found at \url{https://data.openmaterialsdb.se/pressure_induced_invar_effect}.
\section*{Acknowledgements}
We gratefully acknowledge support from the Knut and Alice Wallenberg Foundation (Wallenberg Scholar grant no.
KAW-2018.0194), the Swedish Government Strategic Research Area in Materials
Science on Functional Materials at Link\"oping University (Faculty Grant SFO-Mat-LiU No. 2009-00971) and the Swedish Research Council (VR) Grant No. 2019-05600. B.A. acknowledges financial support from the Swedish Research Council (VR) through Grant No. 2019-05403, as well as support from the Swedish Foundation for Strategic Research (SSF) through the Future Research Leaders 6 program, FFL 15-0290. The computations were enabled by resources provided by the Swedish National Infrastructure for Computing (SNIC) located at National Super Computer Centre (NSC) in Link\"oping, partially funded by the Swedish Research Council through Grant Agreement No. 2018-05973.


%

\end{document}



\title{%
	SUPPLEMENTARY MATERIAL \\
	\large First Principles theory of the Pressure Induced Invar Effect in FeNi Alloys}


\author{Amanda Ehn, Bj\"orn Alling, Igor A. Abrikosov}
\affiliation{Department of Physics, Chemistry, and Biology (IFM), Link\"oping University, SE-58183 Link\"oping, Sweden}




\maketitle

\textbf{Contents: }

\ref{scor}. Spin-pair correlation function trends.

\ref{PBE}. Results with the PBE functional.

\ref{PW91}. Results with the PW91 functional.

\section{Spin-pair correlation function trends}\label{scor}
\begin{figure}[h]
\centering
\includegraphics[width=0.48\textwidth]{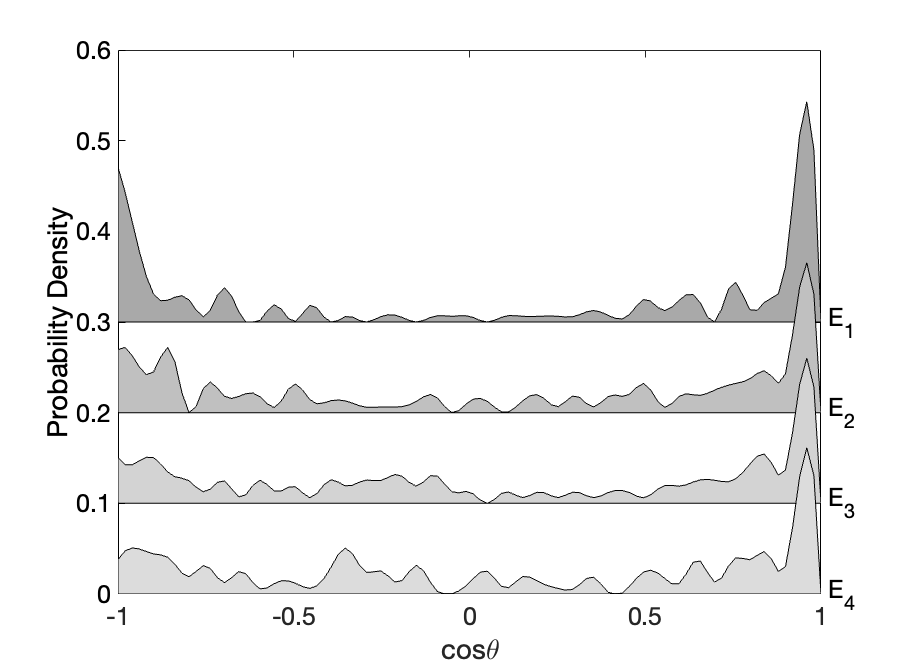}
\caption{\label{sc64}  Spin-pair correlation functions for Fe-Fe NN, from four separate calculations on Fe$_{0.64}$Ni$_{0.36}$ at atomic volume 9.0 \AA$^3$, performed with the LDA functional. The energies are given in Table \ref{t:sc64}, and here $E_1$ has the lowest and $E_4$ the highest energy.}
\end{figure}

\begin{figure}[H]
\centering
\includegraphics[width=0.48\textwidth]{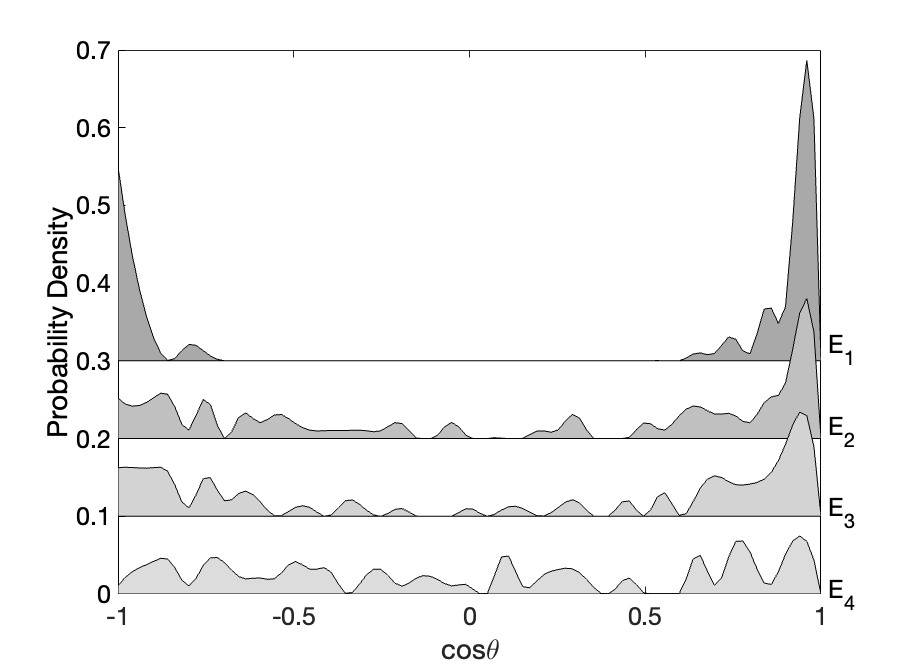}
\caption{\label{sc50}  Spin-pair correlation functions for Fe-Fe NN, from four separate calculations on Fe$_{0.5}$Ni$_{0.5}$ at atomic volume 9.0 \AA$^3$, performed with the LDA functional. The energies are given in Table \ref{t:sc50}, and here $E_1$ has the lowest and $E_4$ the highest energy.}
\end{figure}

\begin{figure}[H]
\centering
\includegraphics[width=0.48\textwidth]{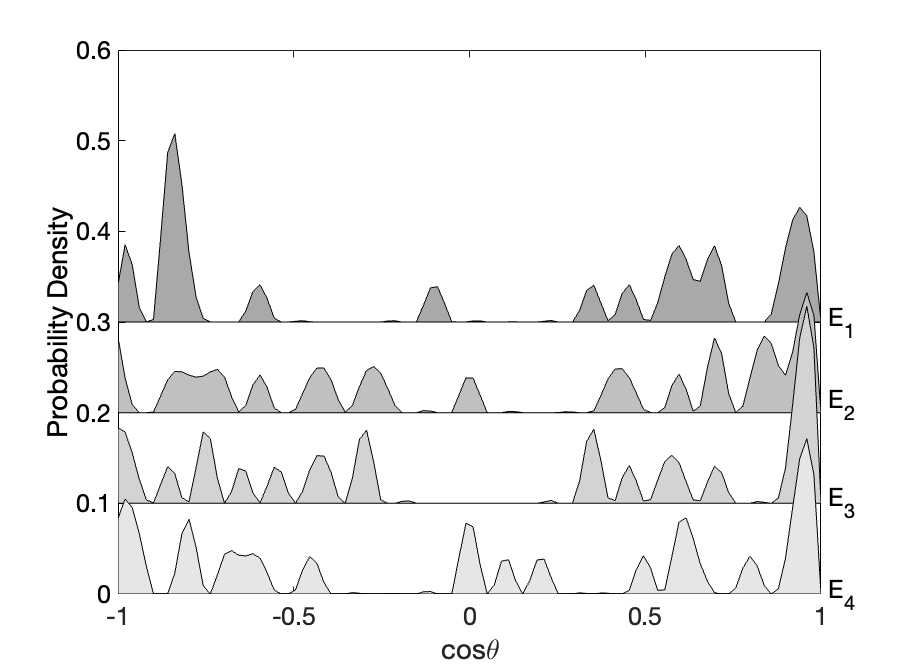}
\caption{\label{sc25}  Spin-pair correlation functions for Fe-Fe NN, from four separate calculations on Fe$_{0.25}$Ni$_{0.75}$ at atomic volume 8.66 \AA$^3$, performed with the LDA functional. The energies are given in Table \ref{t:sc25}, and here $E_1$ has the lowest and $E_4$ the highest energy.}
\end{figure}

\begin{figure}[H]
\centering
\includegraphics[width=0.48\textwidth]{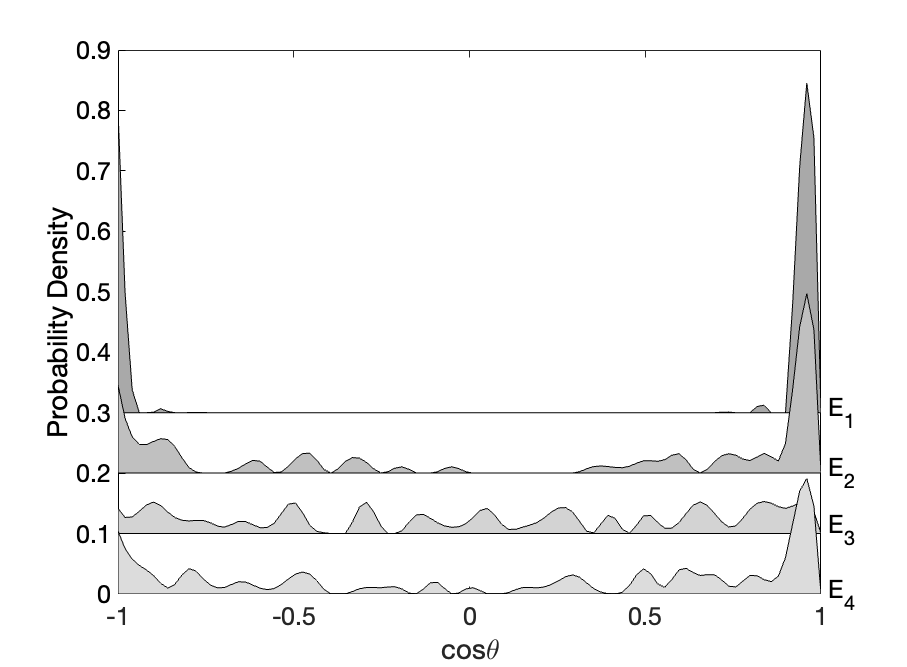}
\caption{\label{scPBE}  Spin-pair correlation functions for Fe-Fe NN, from four separate calculations on Fe$_{0.5}$Ni$_{0.5}$ at atomic volume 9.0 \AA$^3$, performed with the PBE functional. The energies are given in Table \ref{t:scPBE}, and here $E_1$ has the lowest and $E_4$ the highest energy.}
\end{figure}

\begin{figure}[H]
\centering
\includegraphics[width=0.48\textwidth]{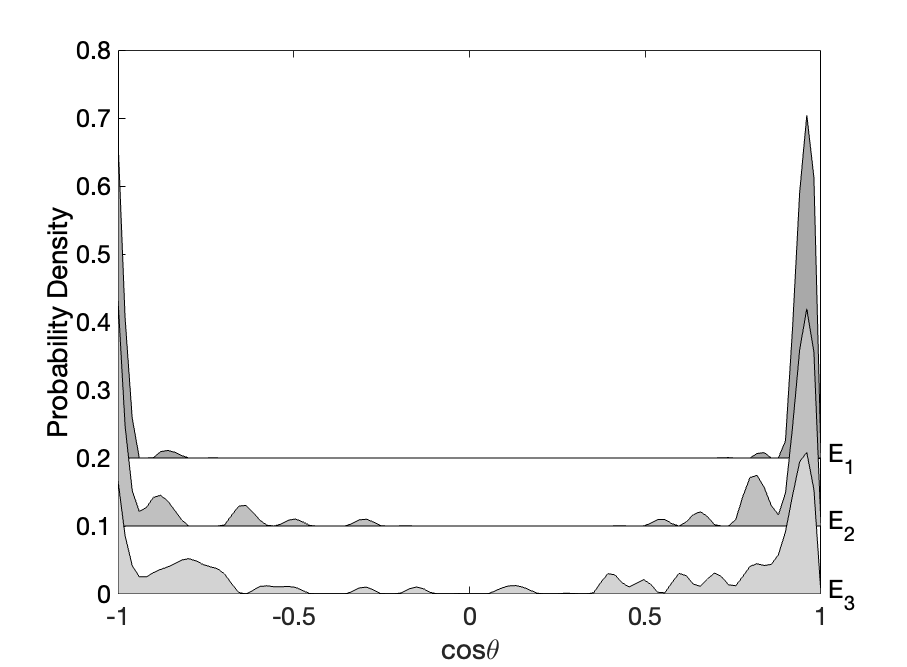}
\caption{\label{scPW91}  Spin-pair correlation functions for Fe-Fe NN, from four separate calculations on Fe$_{0.5}$Ni$_{0.5}$ at atomic volume 9.3 \AA$^3$, performed with the PW91 functional. The energies are given in Table \ref{t:scPW91}, and here $E_1$ has the lowest and $E_3$ the highest energy.}
\end{figure}

For each volume and composition we carry out several calculations with different initial magnetic configurations, leading to different final magnetic states. It is noticed that for the lowest volume for each composition, the energy difference between these configurations are small, as can be seen in Tables \ref{t:sc64}-\ref{t:scPW91} where the energy differences and the average magnetic moment size for Fe atoms are presented. It should be noted that although the average magnetic moment size might be small, generally for the LDA calculations at low volumes, individual moments can still be larger as seen in Figs. 3, 6, and 9 in the main paper and Figs. 8 and 13 here. The magnetic configurations are compared in Figs. \ref{sc64}-\ref{scPW91}, showing the spin-correlation functions for several magnetic configurations at the lowest volume for each composition. 

\begin{table}[h]
    \centering
     \caption{Energy comparison for Fe$_{0.64}$Ni$_{0.36}$ from calculations with LDA for 4 different initial magnetic structures at atomic volume 9.0 \AA$^3$, including average Fe magnetic moment.}
    \begin{ruledtabular}
    \begin{tabular}{cccc} 
   Configuration $i$ &  $E_i$& $\Delta E = E_i - E_1$  & $\langle m_{Fe}\rangle$\\ 
    & (eV/atom) & (meV/atom) & ($\mu_B)$ \\ \hline
     1   & -8.2184 &  0.0 & 0.175 \\ 
     2   & -8.2184 & 0.127 & 0.179\\ 
     3   & -8.2184 &  0.019 & 0.181\\ 
     4   & -8.2184 & 0.019 & 0.181\\ 
    \end{tabular}
    \end{ruledtabular}
   
    \label{t:sc64}
\end{table}

\begin{table}[h]
    \centering
    \caption{Energy comparison for Fe$_{0.5}$Ni$_{0.5}$ from calculations with LDA for 4 different initial magnetic structures at atomic volume 9.0 \AA$^3$, including average Fe magnetic moment.}
    \begin{ruledtabular}
    \begin{tabular}{cccc} 
   Configuration $i$ &  $E_i$& $\Delta E = E_i - E_1$  & $\langle m_{Fe}\rangle$\\ 
    & (eV/atom) & (meV/atom) & ($\mu_B)$ \\ \hline 
     1   & -7.8167 &  0.0 & 0.205\\ 
     2   & -7.8166 & 0.071 & 0.203\\ 
     3   & -7.8166 &  0.0.77 & 0.196\\ 
     4   & -7.8165 & 0.209 & 0.189\\ 
    \end{tabular}
   \end{ruledtabular}
    \label{t:sc50}
\end{table}

\begin{table}[h]
    \centering
    \caption{Energy comparison for Fe$_{0.25}$Ni$_{0.25}$ from calculations with LDA for 4 different initial magnetic structures at atomic volume 8.66 \AA$^3$, including average Fe magnetic moment.}
    \begin{ruledtabular}
    \begin{tabular}{cccc} 
   Configuration $i$ &  $E_i$& $\Delta E = E_i - E_1$  & $\langle m_{Fe}\rangle$\\ 
    & (eV/atom) & (meV/atom) & ($\mu_B)$ \\ \hline 
     1   & -7.0340 &  0.0 & 0.879 \\ 
     2   & -7.0340 & 0.0005 & 0.907\\ 
     3   & -7.0339 &  0.019 & 0.912\\ 
     4   & -7.0316 & 2.351 & 0.854 \\
    \end{tabular}
    \end{ruledtabular}
    
    \label{t:sc25}
\end{table}

\begin{table}[H]
    \centering
    \caption{Energy comparison for Fe$_{0.5}$Ni$_{0.5}$ from calculations with PBE for 4 different initial magnetic structures at atomic volume 9.0 \AA$^3$, including average Fe magnetic moment.}
    \begin{ruledtabular}
    \begin{tabular}{cccc} 
   Configuration $i$ &  $E_i$& $\Delta E = E_i - E_1$  & $\langle m_{Fe}\rangle$\\ 
    & (eV/atom) & (meV/atom) & ($\mu_B)$ \\ \hline 
     1   & -6.567 &  0.0 & 1.086\\ 
     2   & -6.5656 & 1.380 & 1.088\\ 
     3   & -6.565 &  1.970 & 1.114\\ 
     4   & -6.5637 & 3.351 & 0.984 \\
    \end{tabular}
    \end{ruledtabular}
    
    \label{t:scPBE}
\end{table}

\begin{table}[H]
    \centering
    \caption{Energy comparison for Fe$_{0.5}$Ni$_{0.5}$ from calculations with PW91 for 3 different initial magnetic structures at atomic volume 9.3 \AA$^3$, including average Fe magnetic moment.}
    \begin{ruledtabular}
    \begin{tabular}{cccc} 
   Configuration $i$ &  $E_i$& $\Delta E = E_i - E_1$  & $\langle m_{Fe}\rangle$\\ 
    & (eV/atom) & (meV/atom) & ($\mu_B)$ \\ \hline 
     1   & -6.5771 &  0.0 & 1.093\\ 
     2   & -6.5770 & 0.114 & 1.113 \\ 
     3   & -6.5761 &  1.039 & 1.096\\
    \end{tabular}
    \end{ruledtabular}
    
    \label{t:scPW91}
\end{table}

\section{PBE functional}\label{PBE}

\begin{figure}[h]
\centering
\includegraphics[width=.48\textwidth]{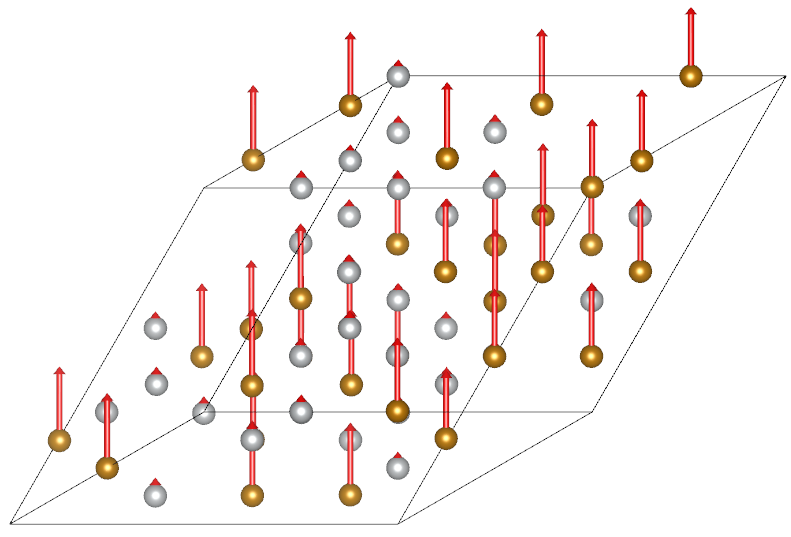}\\
\includegraphics[width=.48\textwidth]{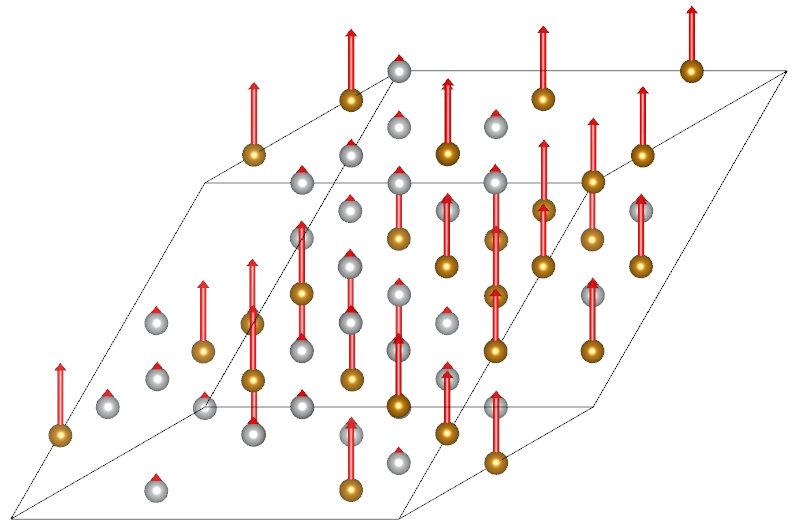}\\
\includegraphics[width=.48\textwidth]{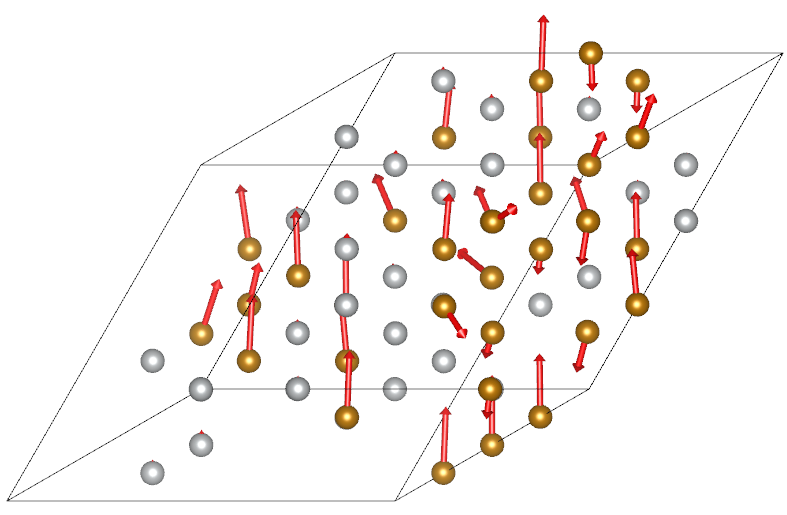}
\caption{Visualisations of the magnetic structure at three volumes for Fe$_{0.5}$Ni$_{0.5}$. From top to bottom: 10.72, 10.18, and 9.65 \AA$^3$. Magnetic transition begins at atomic volume 9.65 \AA$^3$, and the experimental equilibrium volume is 11.45 \AA$^3$. Calculations performed with the PBE functional. Gold and silver spheres represent Fe and Ni, respectively.}
\label{visPBE}
\end{figure}

\begin{figure}[h]
\centering
\includegraphics[width=.48\textwidth]{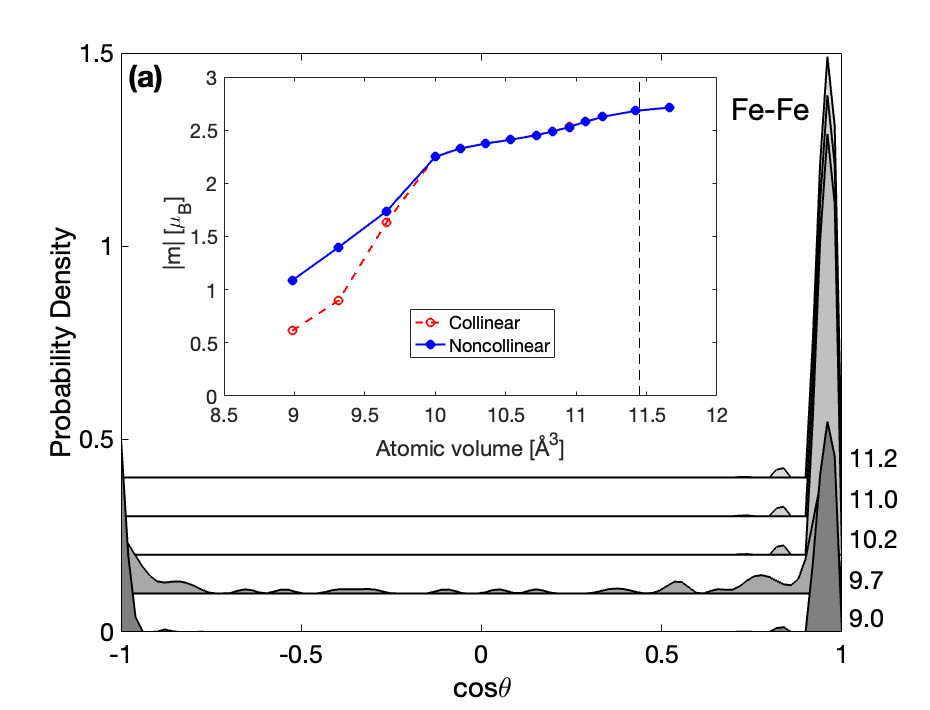}\\
\includegraphics[width=.48\textwidth]{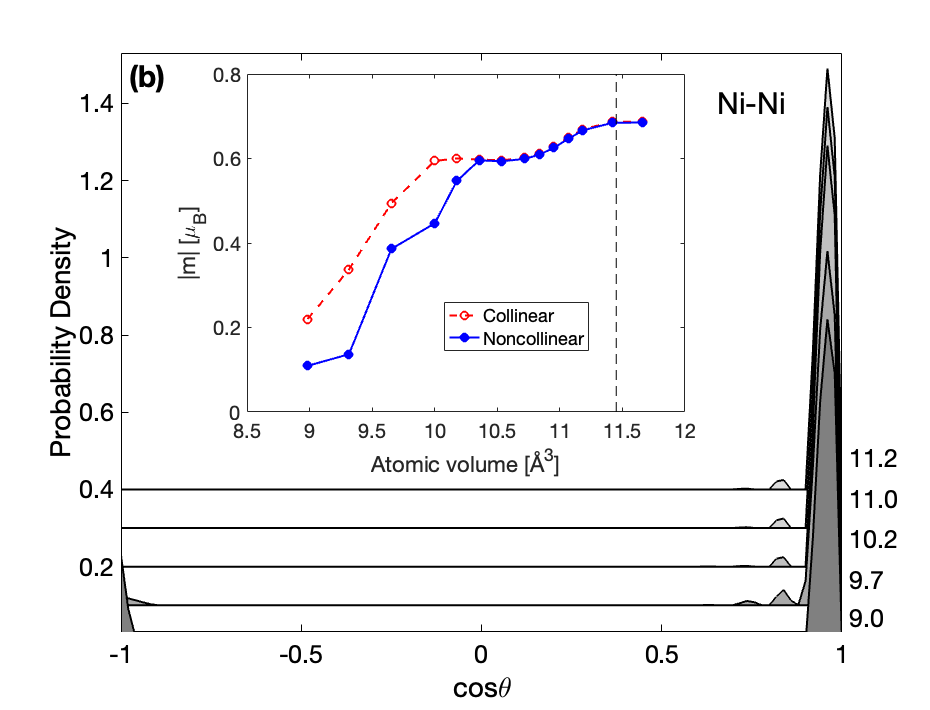}
\caption{Spin-pair correlation functions for Fe$_{0.5}$Ni$_{0.5}$ between neighbouring Fe-spins (panel \textbf{a}) and Ni-spins (panel \textbf{b}) at several volumes (in \AA$^3$) from calculations with PBE. $\hat{s}_i \cdot \hat{s}_j = cos(\theta)$. The right y-axis gives the atomic volumes in units \AA$^3$. Inserted plot shows the average magnetic moment size for noncollinear and collinear calculations in filled in blue circles and red crosses, respectively. Dashed line is experimental equilibrium volume.}\label{two}
\end{figure}

Visualisation of the magnetic evolution for calculations with the PBE functional is presented in Fig. \ref{visPBE}, showing the magnetic structure at three volumes. The magnetic structure is ferromagnetic at high volumes and undergoes a transition to a complex noncolliear state at lower volumes and eventually to an AFM state, as seen in Fig. \ref{two}.

The magnetic evolution of Fe$_{0.5}$Ni$_{0.5}$ shows a transition from a ferromagnetic state at higher volumes to a complex state at low volumes and eventually an AFM state. In contrast to calculations with the LDA functional the magnetic transition begins at the same atomic volume when performing the calculations with the PBE functional. In both noncollinear and collinear calculations the magnetic transition begins at 9.65 \AA$^3$. This transition volume is lower compared to the calculations performed with the LDA functional (10.95 and 10.36 \AA$^3$ for NC and collinear calculations, respectively). The Ni magnetic moments remain parallel at lower volumes than the Fe magnetic moments, as seen in Fig. \ref{two} in which the spin-pair correlation functions are shown for Fe and Ni atoms, respectively. This magnetic evolution is consistent with previous studies that have shown the PBE functional overstabilises the ferromagnetic state, compared to calculations performed with the LDA functional. 

In the equation of state (Fig. \ref{PBEPV}) we observe an anomaly at the volume corresponding to the onset of the magnetic transition. The anomaly is poorly reproduced with regards to pressure, but the amplitude of the anomaly is better reproduced.  The anomaly appears at a volume of 9.65 \AA$^3$, which is 16 \% smaller compared to the experimental equilibrium volume of 11.45 \AA$^3$. As the magnetic transition begins at the same volume in noncollinear and collinear calculations the anomaly is nearly identical in the equation of state based on the two types of calculations. 

\begin{figure}[h]
\centering
\includegraphics[width=0.48\textwidth]{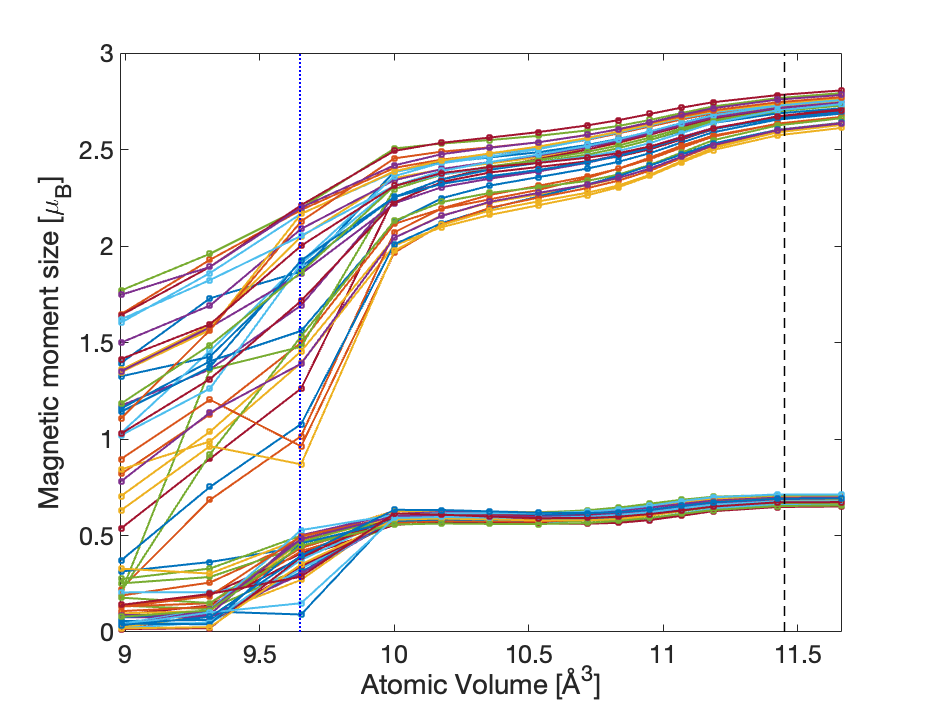}
\caption{\label{PBEMAG}  Magnetic moment sizes for all individual atoms in Fe$_{0.5}$Ni$_{0.5}$, both Fe and Ni, as a function of atomic volume \AA$^3$. Dashed and dotted lines are experimental equilibrium volume and transition volume, respectively.  From calculations with the PBE functional.}
\end{figure}

\begin{figure}[H]
\centering
\includegraphics[width=0.48\textwidth]{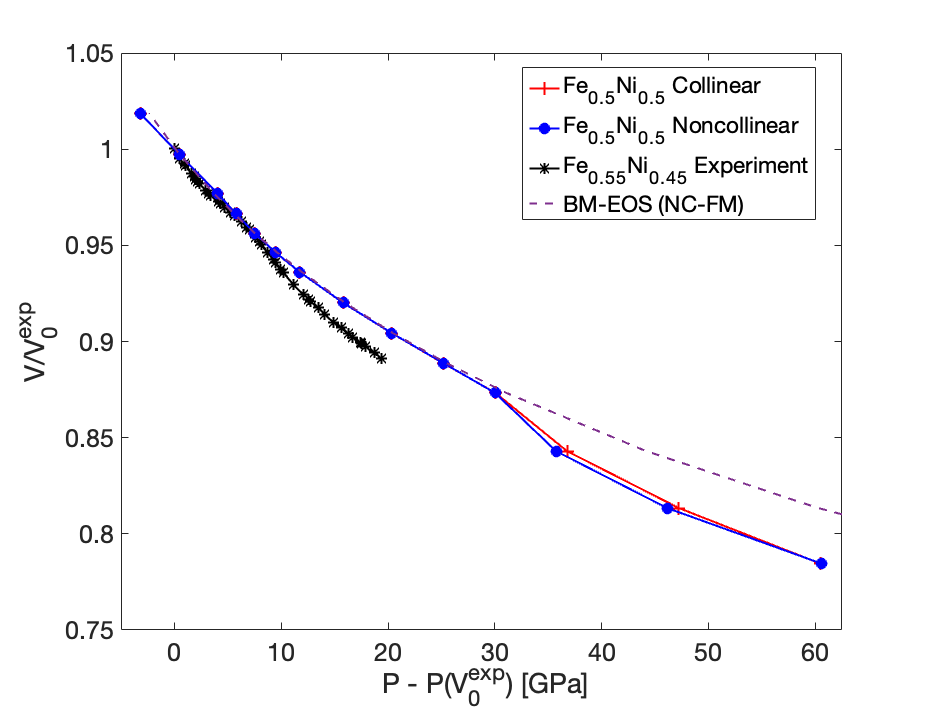}
\caption{\label{PBEPV} Equation of state for Fe$_{0.5}$Ni$_{0.5}$ from NC calculations, collinear calculations, and experiment on Fe$_{0.55}$Ni$_{0.45}$  from \cite{PressureInd}. The dashed line shows the Birch-Murnaghan equation of state fitted to the ferromagnetic datapoints.}
\end{figure}

Further we observe that, in comparison to calculations with the LDA functional, that individual Fe magnetic moments can retain much larger sizes at lower volumes, see Fig. \ref{PBEMAG}. This is also seen in the insets in Fig. \ref{two} where the average magnetic moment size for Fe and Ni atoms are shown, for both NC and collinear calculations. 

\begin{figure}[H]
\centering
\includegraphics[width=0.48\textwidth]{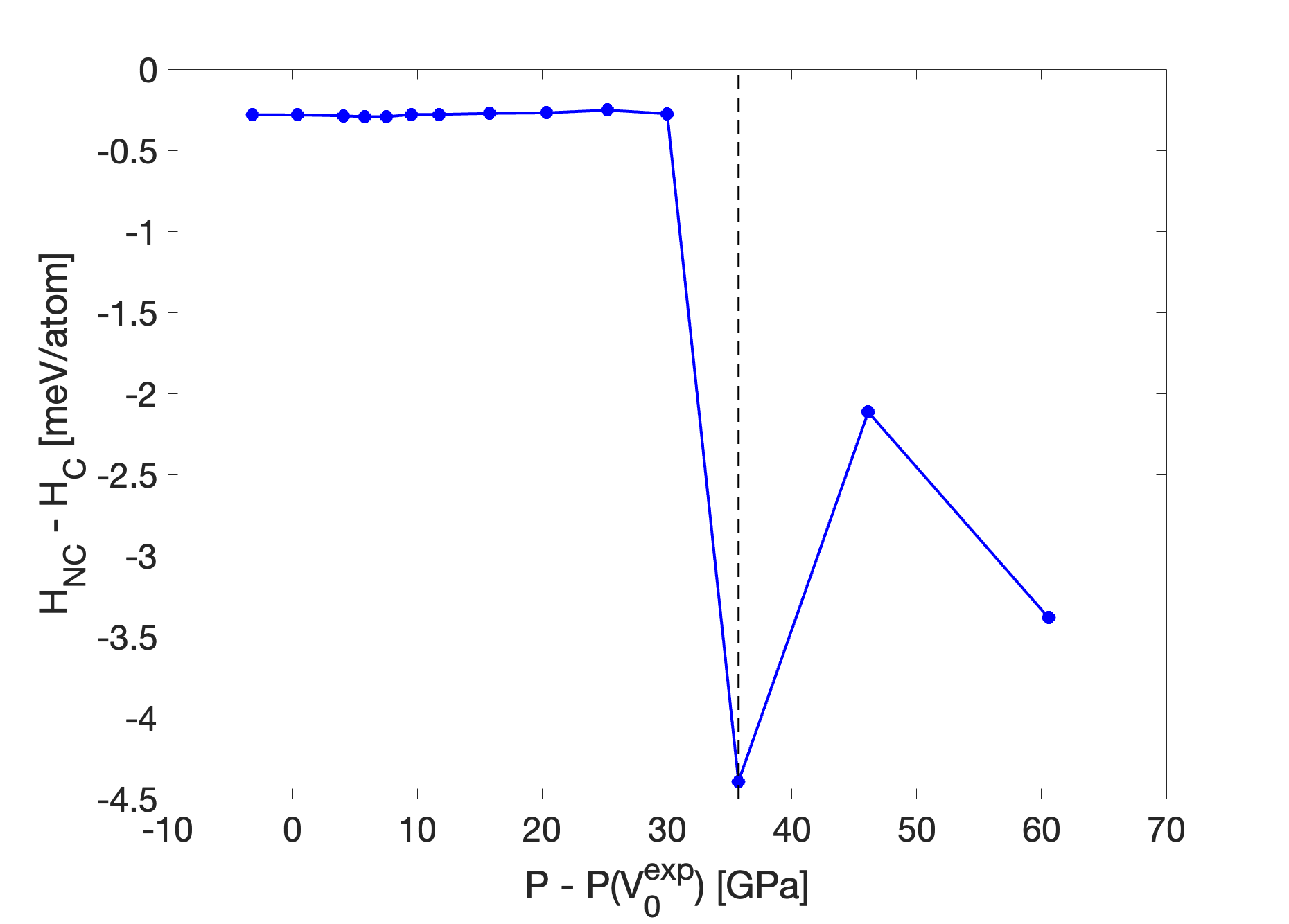}
\caption{\label{PBEen} Enthalpy difference between noncollinear and collinear calculations for Fe$_{0.5}$Ni$_{0.5}$ calculated with the PBE functional. Dashed line represents the magnetic transition in NC calculations.}
\end{figure}

Comparing enthalpies between collinear and noncollinear calculations reveal that the noncollinear results are always favourable, see Fig. \ref{PBEen}. The enthalpies differ most around where the magnetic transition begins. Although the transition occurs at the same volume the magnetic structures are nevertheless different. Noncollinear calculations converge to a magnetic structure with complex state whereas in the collinear case the structure is restricted to spin up and down for individual moments. In the collinear calculations the transition is initiated with a single spin-flip.

It should be noted that the enthalpy difference between NC and collinear calculations should be 0 for the FM solutions. However, we see a nearly constant shift for all compositions and functionals leading to a negative difference for FM solutions. The settings in the calculations are identical apart from allowing noncollinear and collinear magnetic moments, so this seems to be result from the technical settings in VASP.

\begin{table}[h]
 \caption{\label{vol-proc2} Magnetic transition volumes from calculations and $V/V_0^{exp}$ for the studied compositions of fcc Fe-Ni alloys. Experimental equilibrium volume from Ref. \cite{experiment}.} \begin{ruledtabular}
 \begin{tabular}{cccc}
 Composition & $V_0^{exp}$  (\AA$^3$) & $V_{FM \rightarrow NC}$ (\AA$^3$) & $V_{FM \rightarrow NC}/V_0^{exp}$\\ \hline 
 Fe$_{0.5}$Ni$_{0.5}$ & 11.45 & 9.65 & 0.8428 \\
 \end{tabular}
 \end{ruledtabular}
 \end{table}

\section{PW91 functional}\label{PW91}

\begin{figure}[h]
\centering
\includegraphics[width=.48\textwidth]{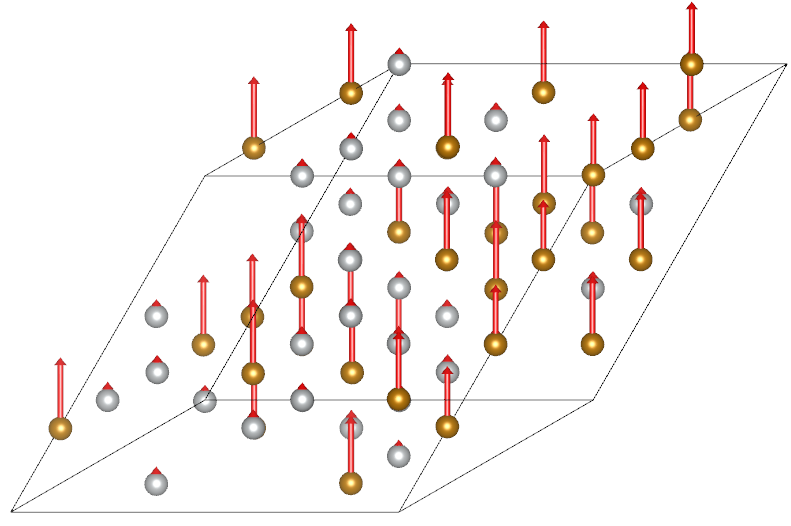}\\
\includegraphics[width=.48\textwidth]{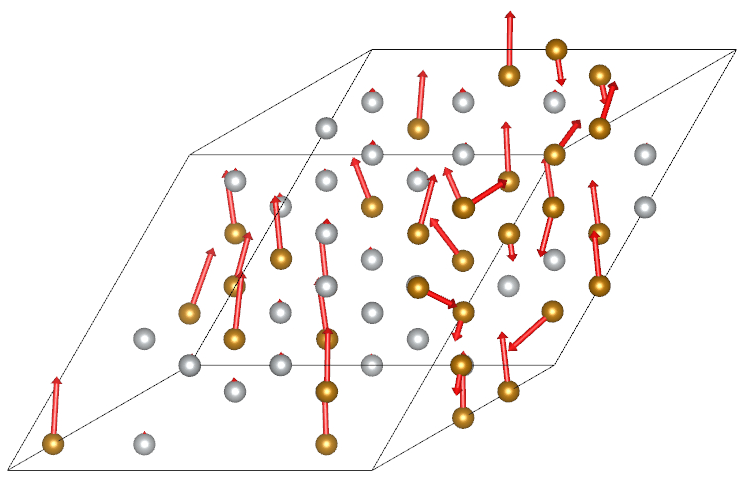}\\
\includegraphics[width=.48\textwidth]{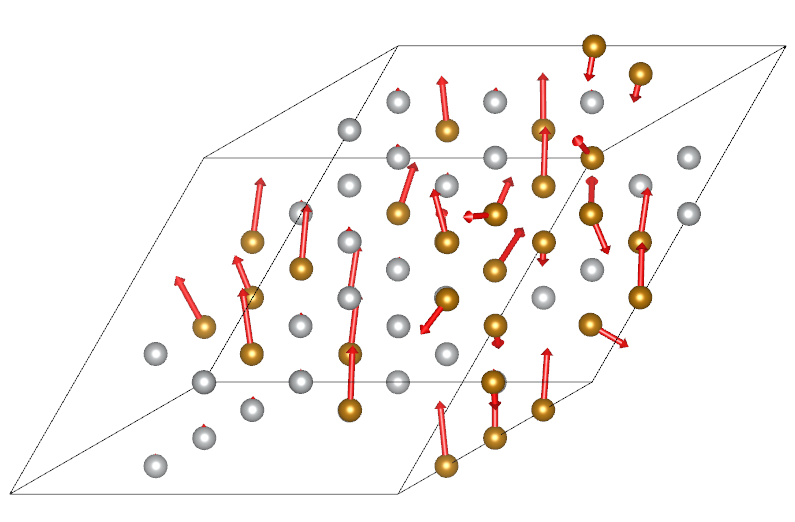}
\caption{Visualisations of the magnetic structure at three volumes for Fe$_{0.5}$Ni$_{0.5}$ with the PW91 functional. From top to bottom: 10.36, 10.18, and 10.00 \AA$^3$. Magnetic transition begins at atomic volume 10.18 \AA$^3$, and the experimental equilibrium volume is 11.45 \AA$^3$. Gold and silver spheres represent Fe and Ni, respectively.}
\label{vispw91} 
\end{figure}

We observe a magnetic transition from ferromagnetic ordering to a complex state also for calculations performed with the PW91 functional for Fe$_{0.5}$Ni$_{0.5}$. The magnetic transition occurs at volume 10.18 \AA$^3$ for noncollinear calculations and 10.00 \AA$^3$ for collinear calculations. The magnetic transition first shows as several individual moments adapting noncollinear arrangements, see Fig. \ref{vispw91}. Previous studies done with the PW91 functional found that the transition begins with a single Fe spin-flip in an Fe-rich local environment. This discrepancy to our study could be explained by the lack of a Fe-rich enough local environment. In \cite{Liot09} a single Fe-atom was surrounded by 11 Fe nearest neighbours while in our supercell a single Fe-atom had at most 9 Fe nearest neighbours.

\begin{figure}[H]
\centering
\includegraphics[width=.48\textwidth]{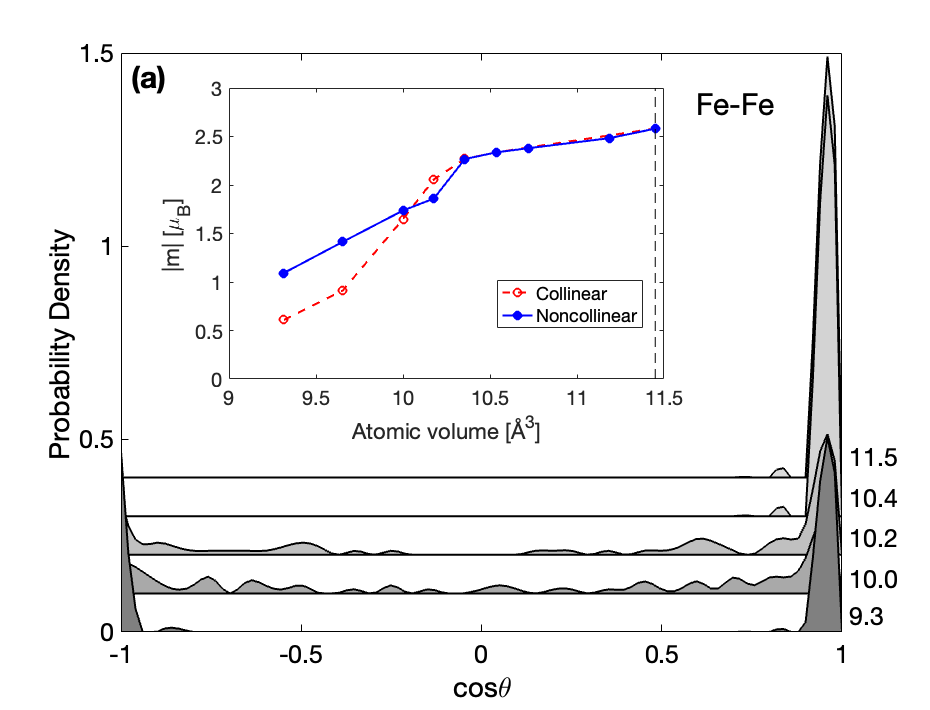}\\
\includegraphics[width=.48\textwidth]{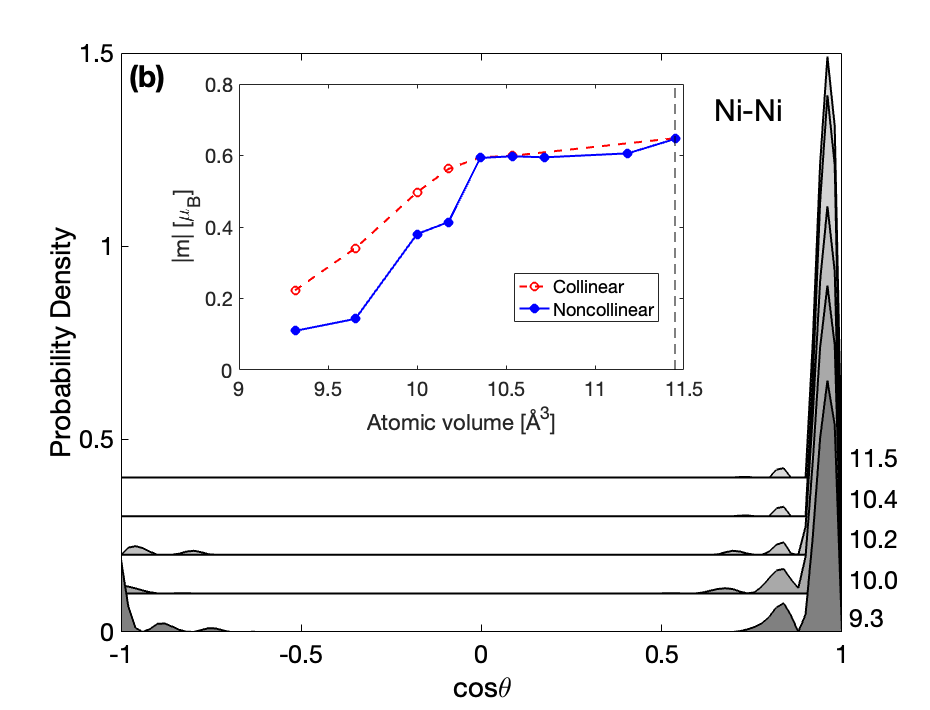}
\caption{\label{PW91_spin}Spin-pair correlation functions between neighbouring Fe-spins (panel \textbf{a}) and Ni-spins (panel \textbf{b}) at several volumes (in \AA$^3$) in Fe$_{0.5}$Ni$_{0.5}$. $\hat{s}_i \cdot \hat{s}_j = cos(\theta)$. The right y-axis gives the atomic volumes in units \AA$^3$. Inserted plot shows the average magnetic moment size for noncollinear and collinear calculations in filled in blue circles and red crosses, respectively. Dashed line is experimental equilibrium volume. Calculations performed with the PW91 functional.}
\end{figure}

The magnetic transition into noncollinear alignments is observed to occur for the Fe-atoms first, as seen in Fig. \ref{PW91_spin}a. The spin-pair correlation function for the Ni-moment reveals that some individual moments flips to anti-parallell alignments at the same volume. Visualisations of the magnetic structure can be seen in Fig. \ref{vispw91} where the magnetic structures at the volumes are shown. Figure \ref{PW91_M} presents the magnetic moment sizes for individual atoms. As with the PBE functional we observe that the magnetic moment sizes of individual moments are larger than compared to LDA calculations, even at low volumes. 

\begin{figure}[H]
\centering
\includegraphics[width=0.48\textwidth]{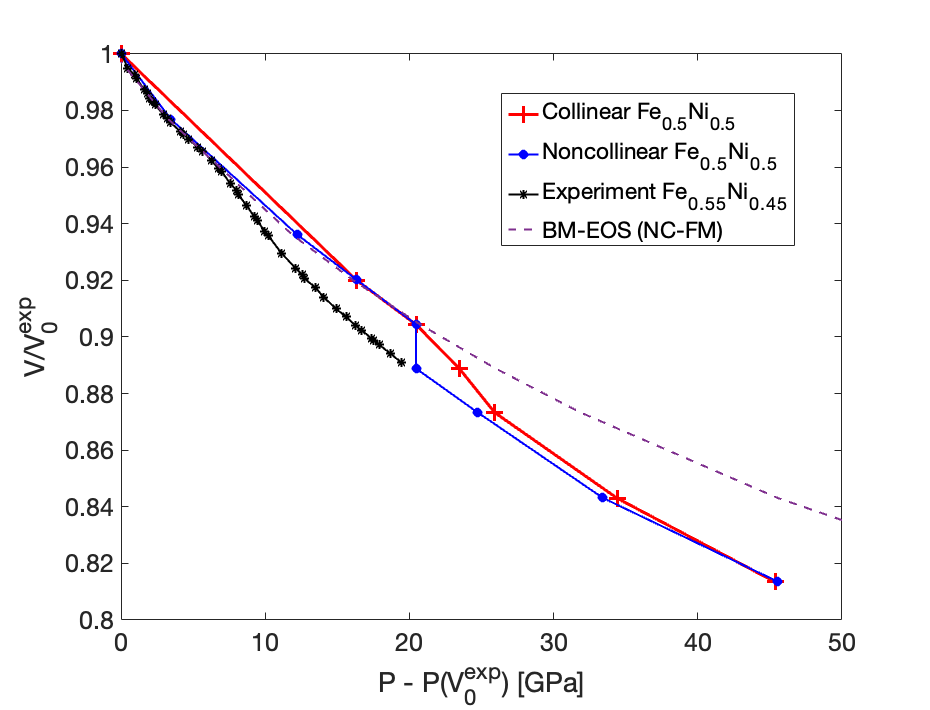}
\caption{\label{PW91PV} Equation of state for Fe$_{0.5}$Ni$_{0.5}$ from NC calculations, collinear calculations, and experiment on Fe$_{0.55}$Ni$_{0.45}$  from \cite{PressureInd}. The dashed line shows the Birch-Murnaghan equation of state fitted to the ferromagnetic datapoints.}
\end{figure}

\begin{figure}[H]
\centering
\includegraphics[width=.48\textwidth]{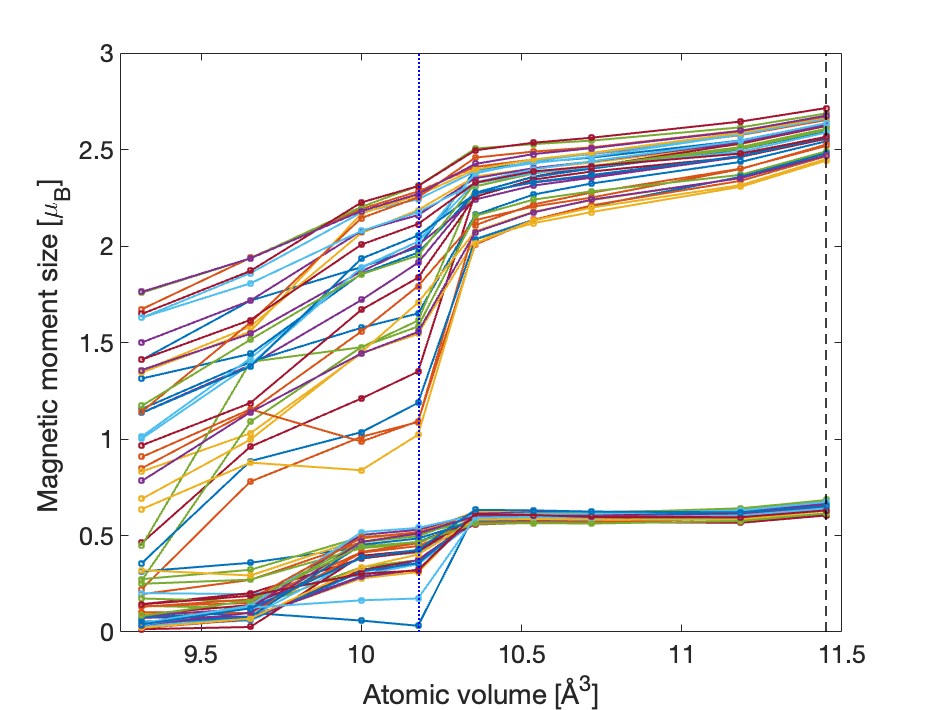}
\caption{\label{PW91_M} Magnetic moment sizes for all individual atoms in Fe$_{0.5}$Ni$_{0.5}$, both Fe and Ni, as a function of atomic volume \AA$^3$. Dashed and dotted lines are experimental equilibrium volume and transition volume, respectively. Calculations performed with the PW91 functional.}
\end{figure}

In the equation of state (Fig. \ref{PW91PV}) we observe an anomaly at the atomic volume where the magnetic transition begins. The anomaly is poorly reproduced in position compared to the anomaly observed in experiments. The PW91 functional reproduces the amplitude of the anomaly best, compared to other functionals used in our study.  The anomaly appears at an atomic volume of 10.18 \AA$^3$, which is 11 \% smaller compared to the experimental equilibrium volume of 11.45 \AA$^3$. An anomaly is seen for the collinear calculations, but it is induced at lower volumes, but once this transition occurs the pressure is nearly identical for collinear and noncollinear calculations. This is also seen in the insets in Fig \ref{PW91_spin} with the average magnetic moment sizes for Fe and Ni moments.

\begin{figure}[H]
\centering
\includegraphics[width=0.48\textwidth]{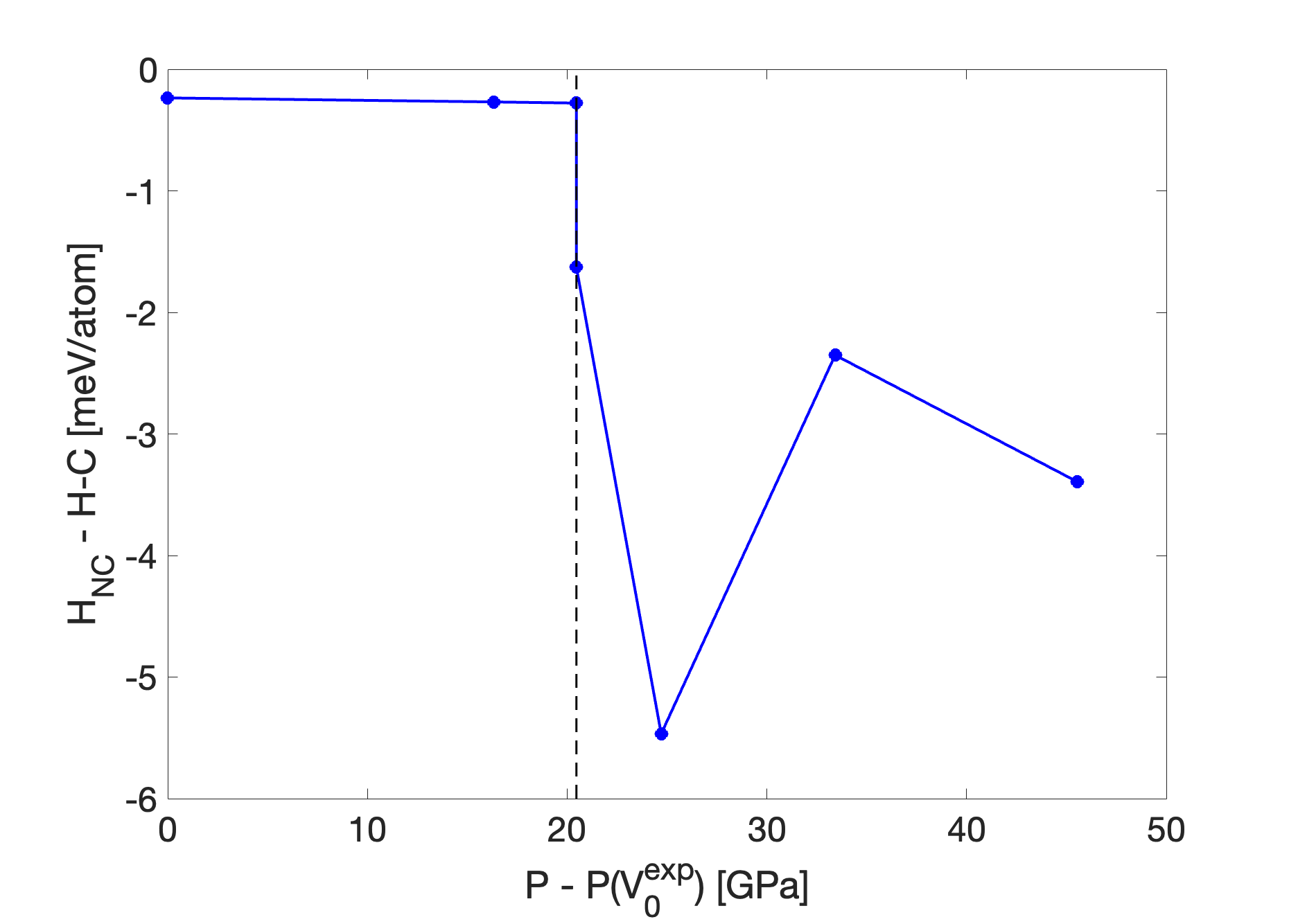}
\caption{\label{P91H} Enthalpy difference between noncollinear and collinear calculations for Fe$_{0.5}$Ni$_{0.5}$ calculated with the PW91 functional. Dashed line represents the magnetic transition in NC calculations.}
\end{figure}

Lastly we compare the enthalpies (Fig. \ref{P91H}) between collinear and noncollinear calculations. We observe that the enthalpy is always lower for the noncollinear calculations. Once the magnetic transition occurs to noncollinear alignments this difference is increased. Once the transition happens for collinear calculations, at higher pressures, this difference is again reduced.

It should be noted that the enthalpy difference between NC and collinear calculations should be 0 for the FM solutions. However, we see a nearly constant shift for all compositions and functionals leading to a negative difference for FM solutions. The settings in the calculations are identical apart from allowing noncollinear and collinear magnetic moments, so this seems to be result from the technical settings in VASP.

\begin{table}[H]
 \caption{\label{vol-proc3}Magnetic transition volumes from calculations and $V/V_0^{exp}$ for the studied compositions of fcc Fe-Ni alloys. Experimental equilibrium volume from Ref. \cite{experiment}.}
 \begin{ruledtabular}
 \begin{tabular}{cccc}
 Composition & $V_0^{exp}$ (\AA$^3$) & $V_{FM \rightarrow NC}$ (\AA$^3$) & $V_{FM \rightarrow NC}/V_0^{exp}$\\ \hline 
 Fe$_{0.5}$Ni$_{0.5}$ & 11.45 & 10.18 & 0.889 \\
 \end{tabular}
 \end{ruledtabular}
 \end{table}


%